\documentclass[journal]{IEEEtran}
\usepackage{cite}
\usepackage{amsmath,amssymb,amsfonts}
\usepackage{mathtools}
\usepackage[table]{xcolor}
\usepackage{algorithmic}
\usepackage{graphicx}
\usepackage{textcomp}
\usepackage{wrapfig} 
\usepackage{orcidlink}
\usepackage[table,xcdraw,x11names]{xcolor}
\usepackage{float} 
\usepackage{subfig}
\usepackage{gensymb}
\usepackage{comment}
\usepackage{soul}
\usepackage[ruled,vlined]{algorithm2e}
\usepackage{multirow}
\usepackage{booktabs}
\usepackage{lipsum}
\usepackage{multicol}
\usepackage{subfiles}
\usepackage{soul}
\usepackage{caption}
\usepackage{subcaption}
\usepackage{enumitem}
\usepackage[dvipsnames]{xcolor}
\setlist[enumerate]{itemsep=0pt}

\usepackage{titlesec}
\titlespacing*{\section}{0pt}{0.4\baselineskip}
{0.4\baselineskip}
\titlespacing{\section}{0pt}{0.4\baselineskip}
{0.4\baselineskip}
\titlespacing*{\subsection}{0pt}{0.4\baselineskip}
{0.4\baselineskip}
\titlespacing{\subsection}{0pt}{0.4\baselineskip}
{0.4\baselineskip}

\definecolor{Gray}{gray}{0.9}
\definecolor{abstractbg}{rgb}{0.89804,0.94510,0.83137}

\def\BibTeX{{\rm B\kern-.05em{\sc i\kern-.025em b}\kern-.08em
    T\kern-.1667em\lower.7ex\hbox{E}\kern-.125emX}}

\makeatletter
\def\ps@IEEEtitlepagestyle{%
  \def\@oddfoot{\mycopyrightnotice}%
  \def\@oddhead{\hbox{}\@IEEEheaderstyle\leftmark\hfil\thepage}\relax
  \def\@evenhead{\@IEEEheaderstyle\thepage\hfil\leftmark\hbox{}}\relax
  \def\@evenfoot{}%
}
\def\mycopyrightnotice{%
  \begin{minipage}{\textwidth}
  \centering \scriptsize
  \copyright 2024 IEEE.  Personal use of this material is permitted.  Permission from IEEE must be obtained for all other uses, in any current or future media, including reprinting/republishing this material for advertising or promotional purposes, creating new collective works, for resale or redistribution to servers or lists, or reuse of any copyrighted component of this work in other works. Published in IEEE Sensors Journal. Final version available at: https://doi.org/10.1109/JSEN.2024.3507951
  \end{minipage}
}
\makeatother

\begin{document}

\makeatletter
\def\ps@IEEEtitlepagestyle{%
  \def\@oddhead{}%
  \def\@evenhead{}%
  \def\@oddfoot{\mycopyrightnotice\hfill\thepage}%
  \def\@evenfoot{\hfill\thepage}%
}
\makeatother

\makeatletter
\def\ps@headings{%
  \def\@oddhead{}
  \def\@evenhead{}%
  \def\@oddfoot{\hfill\thepage\hfill}
  \def\@evenfoot{\hfill\thepage\hfill}%
}
\makeatother

\pagestyle{headings}

\title{Scalable Multi-Subject Vital Sign Monitoring with mmWave FMCW Radar and FPGA Prototyping}

 \author{Jewel Benny$\textsuperscript{a}$, Narahari N. Moudhgalya$\textsuperscript{a*}$, Mujeev Khan$\textsuperscript{b*}$, \\ Hemant Kumar Meena$\textsuperscript{c}$,
Mohd Wajid$\textsuperscript{b}$, Abhishek Srivastava$\textsuperscript{a}$ 
 \\
 \textit{$\textsuperscript{a}$Centre for VLSI and Embedded Systems Technology (CVEST), IIIT Hyderabad, India}\\
 \textit{$\textsuperscript{b}$Department of Electronics Engineering, Z.H.C.E.T. Aligarh Muslim University, India}\\
 \textit{$\textsuperscript{c}$Malaviya National Institute of Technology Jaipur, India}\\
 {$\textsuperscript{*}$These authors have equal contribution to this work}
 
\thanks{ 
This work was supported by the Chips to Startup (C2S) program, Ministry of Electronics and Information Technology (MeitY), Govt. of India, IHub Mobility, IIIT Hyderabad, Kohli Center on Intelligent Systems (KCIS), IIIT Hyderabad and IHub Anubhuti-IIIT Delhi Foundation.
}
}

\maketitle


\begin{abstract}

In this work, we introduce an innovative approach to 
estimate the vital signs of multiple human subjects simultaneously in a non-contact way using a Frequency Modulated Continuous Wave (FMCW) radar-based system.
This work also explores the ambitious goal of extending this capability to an arbitrary number of subjects and details the associated challenges, encompassing both hardware and theoretical limitations. Supported by rigorous experimental results and discussions, the paper paints a vivid picture of the system's potential to redefine vital sign monitoring. An FPGA-based implementation is also presented as proof of concept of an entirely hardware-based and portable solution to vitals monitoring, which improves upon previous works in a multitude of ways, offering 2.7x faster execution and 18.4\% lesser Look-Up Table (LUT) utilization and providing over 7400x acceleration compared to its software counterpart.

\end{abstract}




\vspace{5pt}
\includegraphics[width=0.45\textwidth]{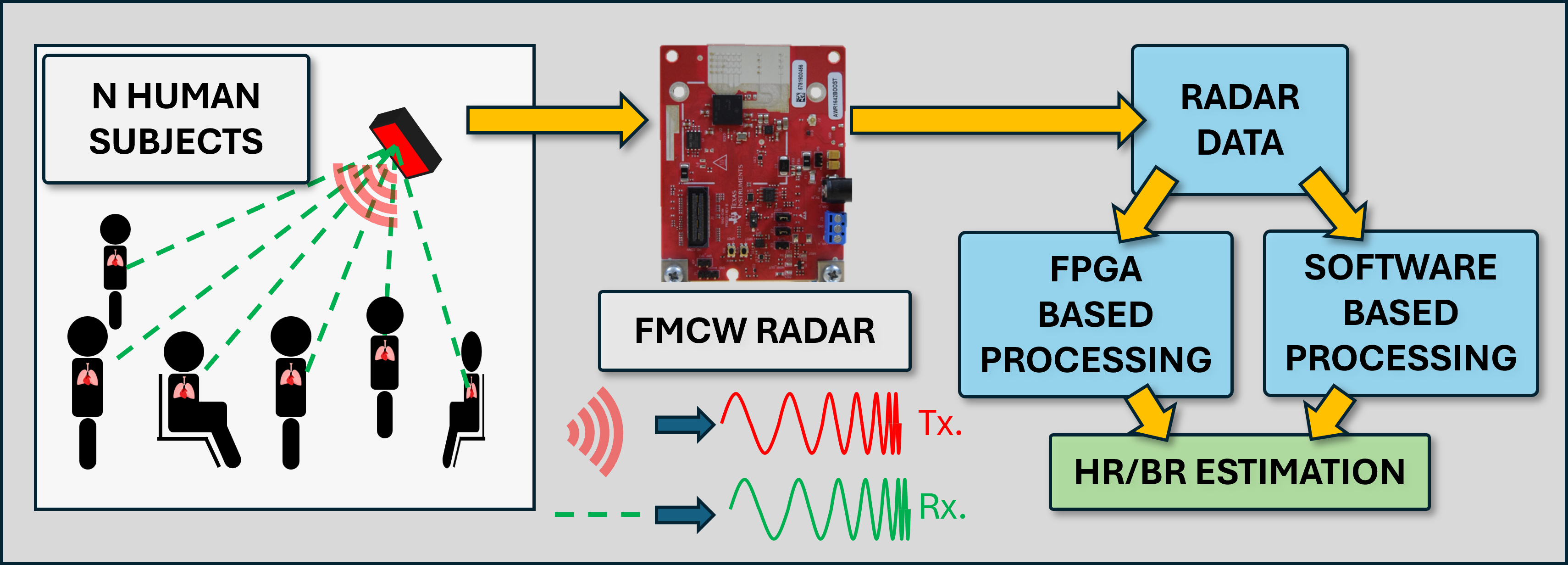}%
\section{Introduction} \label{sec:Introduction}

\IEEEPARstart{R}{ecent} advancements in biomedical and bioelectric engineering have broadened the scope of monitoring human vitals such as Heart Rate (HR) and Breath Rate (BR) through direct or indirect methods utilizing algorithms that analyze data from photoplethysmography (PPG) \cite{PPG}, electrocardiography (ECG) \cite{ECG}, respiratory inductance plethysmography (RIP) \cite{RIP}, and CO\textsubscript{2} measurement (capnography) \cite{capnography}. However, these methods encounter limitations such as subject discomfort with wearables and calibration challenges. A promising solution to overcome these issues is radar sensing technology for HR and BR measurement, offering non-contact capabilities. This approach also extends to applications including sleep apnea detection \cite{sleep_apnea1}, fall detection \cite{fall_detection1} and patient monitoring \cite{patient_monitoring}.


Continuous-wave (CW) Doppler Radar systems have significantly advanced this field, addressing various technical challenges in HR and BR measurement \cite{cw_radar1} \cite{cw_radar2}. However, the fixed frequency of these radars limits their ability to measure the range of subjects, impacting HR and BR measurement in cluttered environments. In contrast, Ultra-Wideband (UWB) radars excel in accurate range measurement and mitigate the limitations of CW radars \cite{uwb_radar2, uwb_radar3, uwb_radar4}.

However, Frequency-modulated-continuous-wave (FMCW) outperforms UWB radars in SNR and accuracy specifically at higher distances \cite{fmcw_vs_uwb}.
\begin{figure}[!hbt]
    \centering
    \includegraphics[width=1\linewidth]{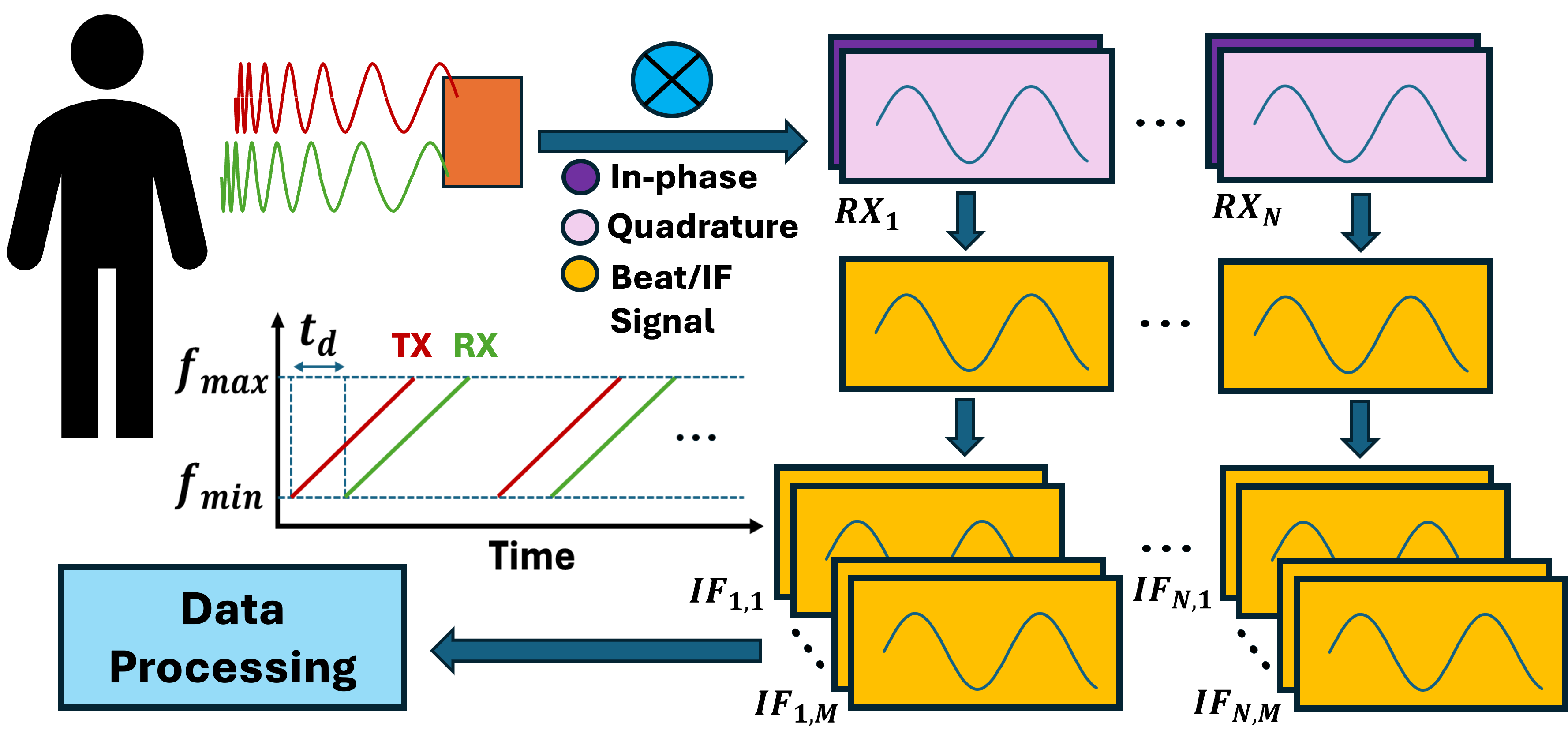}
    \caption{Basic operation principle of FMCW radars for HR/BR measurement}
    \label{basic_principle}
\end{figure}
\textcolor{Black}{
Fig. \ref{basic_principle} depicts the operation principle of FMCW radars. 
FMCW radar continuously transmits a frequency-modulated signal and uses the frequency shift in the reflected signal to measure the target's distance and speed based on delay and Doppler effects \cite{fmcw_radar2, fmcw_radar3, fmcw_radar4, fmcw_radar5, our_sensors_posture, our_iscas_posture, our_icra}.
}
The superior SNR of FMCW radars is particularly beneficial for assessing vital signs, especially in scenarios involving multiple people, as outlined in \cite{fmcw_radar2}. Furthermore, UWB radars suffer from low SNR in HR detection due to spurious harmonics and mixed respiration and heartbeat signals, which can corrupt FFT spectrograms \cite{uwb_radar2}. FMCW radars have several other advantages over UWB radars. While UWB radars are limited by pulse width and peak signal intensity \cite{cw_uwb_fmcw_5, cw_uwb_fmcw_6, cw_uwb_fmcw_7} and require high power during short pulse periods \cite{cw_uwb_fmcw_8}, FMCW wideband radars combine the ranging capability of UWB radars with the sensitivity and robustness of Continuous Wave (CW) Doppler radars. This combination makes FMCW radars particularly effective for detecting micro-motions, such as chest wall oscillations \cite{cw_uwb_fmcw_9}, and monitoring multiple targets at different ranges \cite{cw_uwb_fmcw_10}. Moreover, FMCW radars are typically smaller, lighter, more power-efficient, and support real-time processing \cite{cw_uwb_fmcw_11}, which enhances their reliability in suboptimal environments like hospitals and homecare settings.

\begin{figure}[]
    \centering
    \includegraphics[width=1\linewidth]{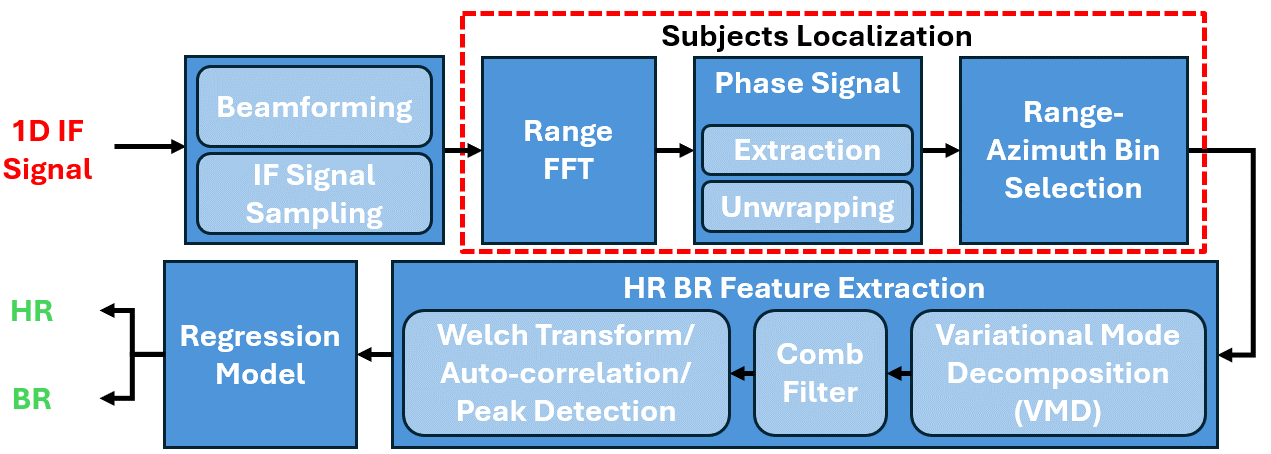}
        \caption{Proposed algorithm integrating VMD, Comb Filter, and Regression Model for improved MAE in vital measurements.}
    \label{fig:proposed_algo}
\end{figure}

Previous works have focused on HR and BR measurements for both single subjects \cite{fmcw_radar1} and multiple subjects \cite{fmcw_radar2, fmcw_radar3, fmcw_radar4, fmcw_radar5}, tackling challenges in signal processing and subject localization. Our previous work \cite{prev_work} achieved improved HR and BR measurement accuracy compared to these works by combining multiple measurements using a least squares approach to improve noise performance. 
The mentioned works position radar at chest level to maximize Doppler information (DI), which refers to the frequency shifts caused by the movement of objects relative to the radar, essential for determining their speed and direction. 
\textcolor{Black}{
However, this setup limits the vital measurements to one subject per azimuth,
as subjects behind others remain obscured from the radar’s field of view.
To overcome this challenge, 
in this work we strategically adjust radar's height and tilt 
to ensure that multiple subjects can be within its 
field of view.
However, these adjustments 
result in an increased Mean Absolute Error (MAE) in vital measurement. 
}
\textcolor{Black}{
Therefore, as depicted in Fig. \ref{fig:proposed_algo}, 
in this work 
we present a unique phase signal extraction algorithm to address the challenge of increased MAE in multi-subject vital measurement scenario. 
As shown in Fig. \ref{fig:proposed_algo}, the proposed algorithm integrates Variational Mode Decomposition (VMD) to filter out noise, and a comb filter to eliminate breathing harmonics, thus enhancing the quality of the heartbeat signal. 
Moreover, to further reduce MAE, features from these refined signals are fed into a regression model, enabling more accurate predictions of HR and BR.
Additionally, we also 
present the  hardware implementation 
of the proposed algorithm on a Field Programmable Gate Array (FPGA), which achieves a significant improvement over prior methods
\cite{tudose2018pulse,tsao2018two,heo2021fpga}.
}
\textcolor{Black}{
Our design offers 2.7x faster execution, reduces Look-Up Table (LUT) utilization by 18.4\%, and accelerates performance by over 7400x compared to its software counterpart.
}

This article is organized as follows - Section \ref{sec:Proposed Algorithm} delves into the fundamentals of FMCW radar signal processing and then outlines the proposed algorithm for extracting HR and BR features from multiple subjects within the radar's field of view. Section \ref{sec:Practical Constraints} discusses the hardware constraints and empirical trade-offs in detecting the vitals for an arbitrary number of subjects. Section \ref{sec:HW Implementation} discusses the system implementation and comparative performance analysis of the proposed algorithm performance, and finally, Section \ref{sec:Conclusion} concludes the article.
\section{Fundamentals of FMCW Radar and Proposed Algorithm for HR and BR Extraction} \label{sec:Proposed Algorithm}
Fig. \ref{basic_principle} depicts the operation principle of FMCW radars. 
The radar sends a chirp signal at light speed $c$, which echoes off objects in its path and is then received by the radar.
For a linear chirp with amplitude $A_{tx}$, initial frequency $f_{min}$, chirp slope $K_c$, round trip time delay for reception $t_d$ and attenuation factor  $\alpha$, the transmitted signal $y_{tx}(t)$ and received signal $y_{rx}(t)$ are defined by Eq. \ref{eq:tx} and \ref{eq:rx}\textcolor{Black}{, respectively}. 
\begin{align}
y_{tx}(t) & = A_{tx}\cos(2\pi f_{min}t + K_c\pi t^2) \label{eq:tx}\\
y_{rx}(t) & = \alpha y_{tx}(t-t_d) \label{eq:rx} 
\end{align}
As shown in Fig. \ref{basic_principle}, a mixer 
combines 
TX and RX chirps from multiple transmitters and receivers, followed by low-pass filtering. The mixer's output, the intermediate frequency (IF) signal, is expressed by Eq. \ref{Eq: 3}. The distance $R$ of the object from the radar is directly related to $t_d$ as $R = \frac{c\times t_d}{2}$. Since $R$ is significantly less than $c$ in magnitude, $t_d$ is typically in microseconds, and hence the higher order term of the IF signal in Eq. \ref{Eq: 3} can be neglected, resulting in Eq. \ref{mixer_out}
\begin{align} 
y_{IF\_LPF}(t) & = \text{LPF}\Big(y_{tx}(t)\times y_{rx}(t)\Big) \label{Eq: 3}\\
\begin{split}
& = A\cos(2\pi f_{min}t_d+2\pi K_ctt_d-\pi K_ct_d^2) \\
\end{split} \notag \\
& \approx A\cos(2\pi f_{min}t_d+2\pi K_ctt_d) \label{mixer_out}  
\end{align}

The radar data is observed over time and digitized for processing. The quadrature phase shifted IF signal is also generated, resulting in the complex IF signal with IF frequency $f_b = K_ct_d$, and can be given by Eq. \ref{IF_signal}. The term $\phi_{R}$ is the phase change associated with an object at range $R$. If $i$ number of objects are present, the transmitted chirp will bounce off each of these, and the resulting IF signal will be the summation of the individual IF signals corresponding to each object given by Eq. \ref{IF_signal_multi}, which forms the basis of our proposed algorithm. 
\begin{align}
    y_{IF\_Quadrature}(t) & = A\exp{\Big(j(\big(2\pi f_{min}t_d+2\pi K_ctt_d\big)\Big)} \notag\\
    & = A\exp{\Big(j\big(2\pi f_bt + \frac{4\pi R}{\lambda_{max}}\big)\Big)} \notag \\
    & = A\exp{\Big(j\big(2\pi f_bt + \phi_R\big)\Big)} \label{IF_signal}
\end{align}
\begin{align}
    y_{IF}(t) & = \sum \limits_i y_{IF\_Quadrature_i}(t) \notag \\
    & = \sum \limits_i A_i\exp{\Big(j\big(2\pi f_{b_i}t + \phi_{R_i}\big)\Big)} \label{IF_signal_multi}
\end{align}
\textcolor{Black}{
Static reflectors within the subject's range or angular bin can affect the Doppler signal and introduce a direct current (DC) offset \cite{dc_2}. This offset merges with the intermediate frequency (IF) signal, as represented in Eq. 6, potentially causing signal distortion and lowering the precision of vital sign measurements. Therefore, correcting the DC offset is an essential pre-processing step in the algorithm. To achieve this, we employed the non-linear least squares (NLLS) estimation technique to minimize the geometric distance \cite{fmcw_radar1}, as it provides the highest accuracy for mmWave radars \cite{dc_2}.
}

Fig. \ref{fig:proposed_algo} depicts the proposed algorithm using a multi-receiver radar for accurate localization and estimation of HR and BR of multiple subjects. The following sections describe each step in the proposed algorithm in detail.

\subsection{IF Signal Sampling for Different Azimuths Using Beamforming} \label{subsec:IF Signal Sampling}
Azimuth information cannot be retained with the one-dimensional IF signal from Eq. \ref{IF_signal_multi}. 
A multi-antenna radar, utilizing multiple TX-RX pairs and beamforming \cite{fmcw_radar2}, produces a two-dimensional IF signal matrix. This setup enables determination of azimuth and range for each subject. The matrix is derived by summing IF signals from all TX-RX pairs ($J$ in total) as given in Eq. \ref{y_IF,k(t)},
\begin{equation}
y_{IF,k}(t) = \sum\limits_{j=1}^J y_{IF}^{j}(t)w_k^j \label{y_IF,k(t)}
\end{equation}
This is a one-dimensional IF signal for each azimuth bin, where $j$ and $k$ correspond to the index of azimuth and antenna pair\textcolor{Black}{, respectively}. 

\subsection{Range FFT to Determine IF Signal Frequency Components}  \label{subsec:Range FFT}
It is crucial to determine the IF signal frequency components for further steps in the algorithm,
and can be computed from its magnitude spectrum for each IF signal sampled at a specific azimuth. As the frequency of the IF signal corresponds to the range of an object, the magnitude spectrum will exhibit peaks at $R_i = \frac{cf_{b_i}}{2K_c}$ for each object situated in front of the radar. The argument for the maximum of magnitude spectrum, $\big|\Big(Y_{IF,k}(f)\Big)\Big|$=$\big|\mathcal{F}\Big(y_{IF,k}(t)\Big)\Big|$, gives the range $R_0^k$ as defined by Eq. \ref{equation9a}, where $y_{IF,k}(t)$ is the IF signal obtained from the $k^{\text{th}}$ azimuth bin and $R_0^k$ is the range of an object present in this azimuth bin. The \textit{Range FFT} can be defined by Eq. \ref{equation9b}. So, the peak samples of \textit{Range FFT} are given by Eq. \ref{equation9}.
\begin{equation} \label{equation9a}
\begin{split}
R_0^k=\big(\frac{c}{2K_c}\big)\times arg\:\Big(max\: \big|\mathcal{F}\Big(y_{IF,k}(t)\Big)\big|\Big)
\end{split}
\end{equation}
\begin{equation} \label{equation9b}
\begin{split}
Y_{k}\big(r) = \mathcal{F}\Big(y_{IF,k}(t)\Big)\big|_{f=\frac{2K_cR}{c}}
\end{split}
\end{equation}
\begin{equation} \label{equation9}
\begin{split}
Y_{k}\big(R_0^k) \approx \mathcal{F}\Big(A\exp(j(2\pi ft))\Big)\Big|_{f=f_b}\exp\Big(\frac{4\pi [R_0^k]}{-j\lambda_{max}}\Big)
\end{split}
\end{equation}

Interestingly, owing to the limited penetration capabilities of chirps, the magnitude of each peak will decrease as the range increases for a specific azimuth. 

\subsection{Phase Signal Extraction From Temporal Range FFT Information} \label{subsec:Phase Signal Extraction}

The \textit{Range FFT} spectrum $Y_{k}$ retains critical phase information linked with the IF signal as seen in Eq. \ref{equation9}. 
Since human chest movement ranges from $\sim$1 to 12 mm for respiration and $\sim$0.01 to 0.5 mm for heartbeat \cite{chest_disp_ref2}, the subject's range information undergoes slight temporal variations, as expressed in Eq. \ref{eq:R(t)}, where $x(t)$ is the chest displacement signal. So, the IF signal of the $k^{\text{th}}$ azimuth bin is expressed as shown in Eq. \ref{eq:IF_kth_Azi}. The temporal phase change is hence given by Eq. \ref{eq:phase_change}.
\begin{equation} \label{eq:R(t)}
    R(t) = R + x(t)
\end{equation} 
\begin{equation} \label{eq:IF_kth_Azi}
\begin{split}
y_{IF,k}(t) & = \sum\limits_{j=1}^J\sum\limits_{i=1}^I A_i^j\exp{\Big(j\big(2\pi f_{b_i}^jt + \phi_{R_i}^j\big)\Big)}w_k^j \\
& =  \sum\limits_{j=1}^J\sum\limits_{i=1}^I A_i^j\exp{\Big(j\big(2\pi f_{b_i}^jt + \frac{4\pi \big(R_i^j+x_i^j(t)\big)}{\lambda_{max}}\big)\Big)}w_k^j
\end{split}
\end{equation}


\begin{equation} \label{eq:phase_change}
    \Delta\phi(t) = \frac{4\pi}{\lambda_{max}}x(t)
\end{equation}

Given the millimeter-scale wavelengths of FMCW radars, even small oscillations in $x(t)$ cause significant phase changes. As $|\Delta\phi(t)| \leq \pi$, displacements exceeding $\frac{\lambda_{max}}{4}$ wrap into the phase. Thus, phase unwrapping is crucial for preserving accurate phase information \cite{phase_unwrap}. 
The true phase can be recovered from the wrapped phase by simply calculating the phase difference between every two consecutive samples in the wrapped phase signal and then adding/subtracting multiples of 2$\pi$  where the phase has been accumulated. 

\subsection{Range-Azimuth Bin Selection for Localizing Subjects} \label{subsec:Range-Azimuth Bin Selection}

The next step is to localize the subjects to extract only the required unwrapped phase signal samples. In range-azimuth bins without objects or with stationary clutter, $\Delta\phi(t)$ is negligible. However, $\Delta\phi(t)$ exhibits high variance in bins with subjects, resulting in significant 
DI 
variance in the \textit{Range FFT}. 
This variance distinctly separates the frequency distribution of DI in azimuth bins with subjects from those lacking them.

\begin{figure}[]
    \centering
    \includegraphics[width=0.85\linewidth]{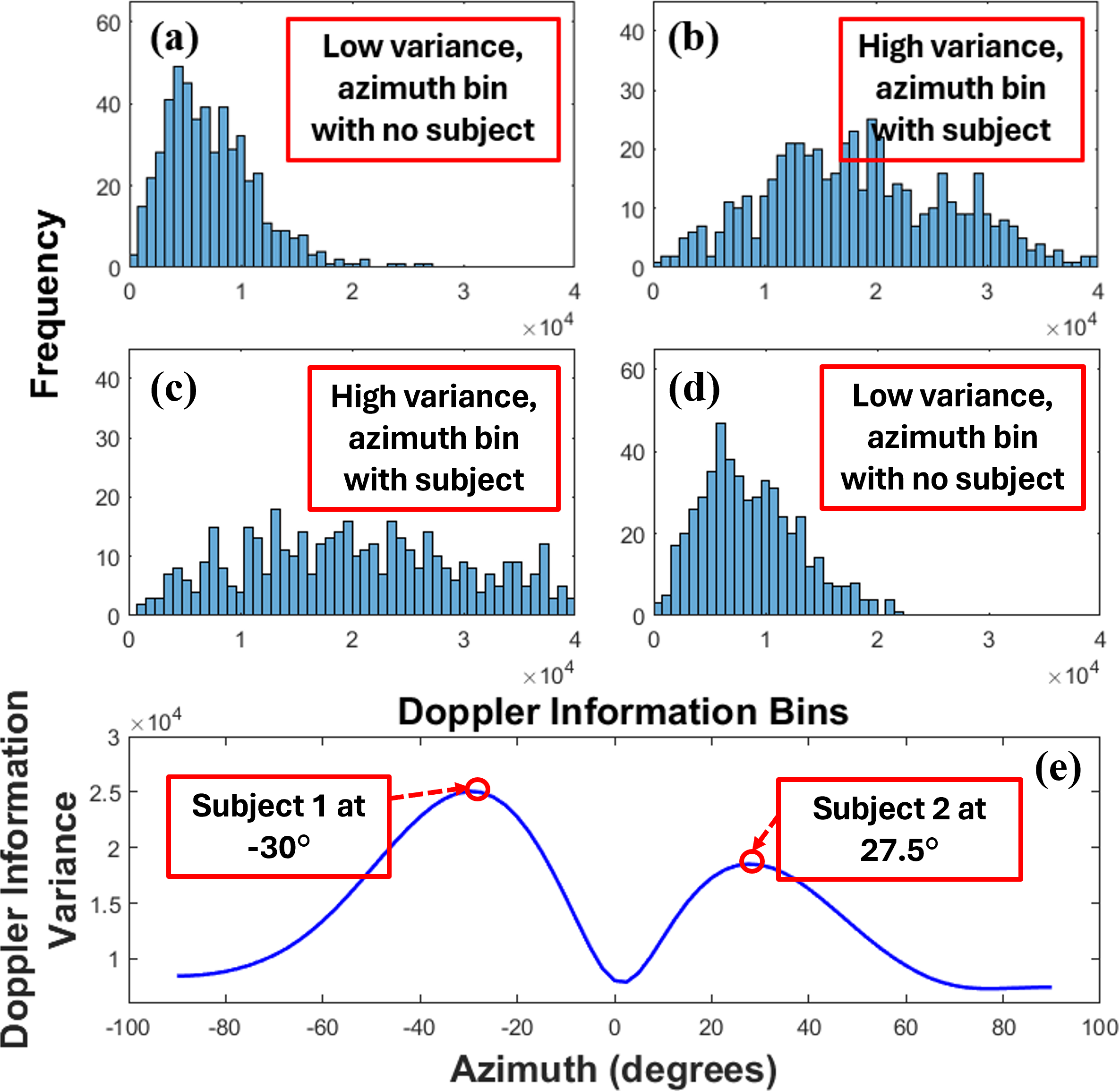}
    \caption{\textcolor{Black}{DI} Spread vs. Azimuth - The plots \textit{(a)}-\textit{(d)} show the distribution of DI amplitude across azimuth bins, and plot \textit{(e)} shows variance vs. azimuth.}
    \label{doppler_info_vs_azimuth}
\end{figure}

\begin{figure}[]
    \centering
    \includegraphics[width=0.85\linewidth]
    {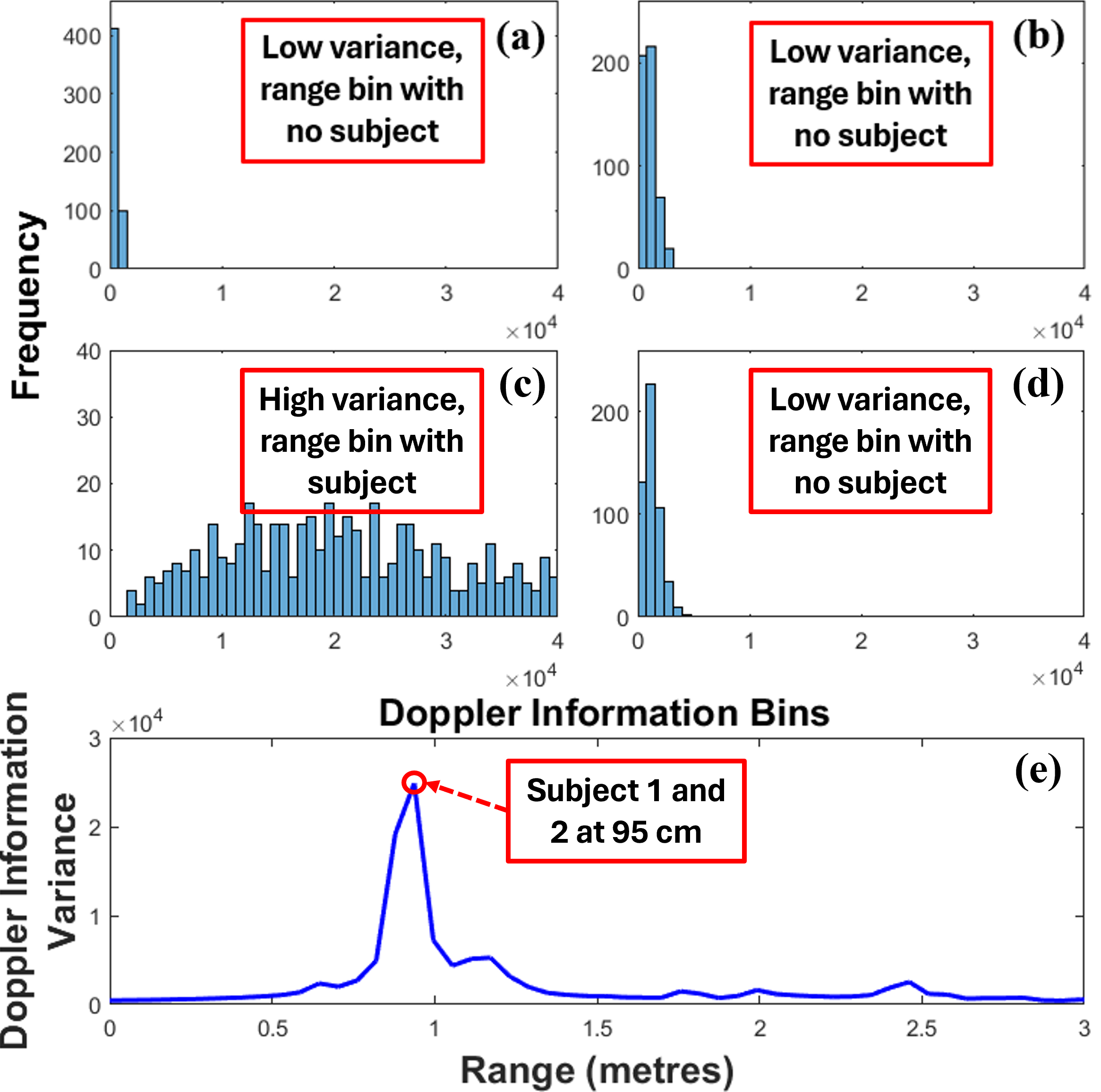}
    \caption{\textcolor{Black}{DI} Spread vs. Range - The plots \textit{(a)}-\textit{(d)} show the distribution of DI amplitude across range bins and plot \textit{(e)} shows variance vs. range.}
    \label{doppler_info_vs_range}
\end{figure}

\begin{figure}[]
    \centering
    \includegraphics[width=0.6\linewidth]
    {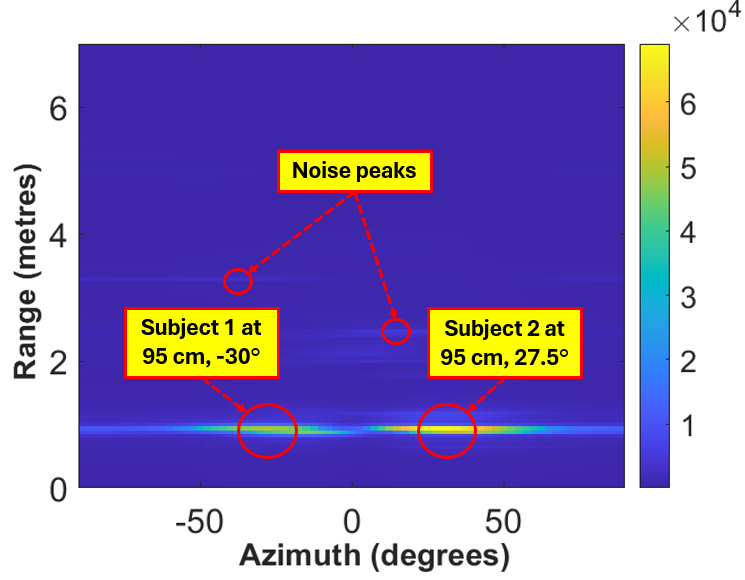}
    \caption{Range-Azimuth map showing DI variance - Two distinct peaks (subjects) are visible along with noise peaks with low DI variance.}
    \label{range_azi_map}
\end{figure}

\textcolor{Black}{
Fig. \ref{doppler_info_vs_azimuth} and \ref{doppler_info_vs_range} show the DI variance plotted against range and azimuth, respectively. We observe that the DI variance \textcolor{Black}{ peaks} at the range and azimuth bins corresponding to the subjects' locations. As a result, localizing the subjects involves identifying these peaks in the DI variance across the range-azimuth space. For each azimuth bin, the Range FFT provides the DI variance for each range bin over time, simplifying the two-dimensional peak-finding problem into several one-dimensional peak-finding tasks. Ideally, the number of peaks should correspond to the number of subjects, provided there are no other significant sources of DI variance in the environment.}
\textcolor{Black}{
In cluttered environments, additional peaks may appear due to environmental noise and other factors. However, the DI associated with these noise peaks is typically much lower than that of human subjects, making it possible to filter them out using simple thresholding techniques. Fig. \ref{range_azi_map} shows the Range-Azimuth map, where two prominent peaks corresponding to the subjects' locations are clearly visible, along with some noise peaks that have much lower DI variance amplitudes.}

\subsection{HR/BR Features Extraction and Measurement} \label{subsec:HR/BR Feature Extraction}
Once the required unwrapped phase signal samples from the range-azimuth bins of the subjects are extracted, the next step involves extracting HR and BR features from these signals. 

\begin{figure}
    \centering
    \includegraphics[width=0.4\textwidth]{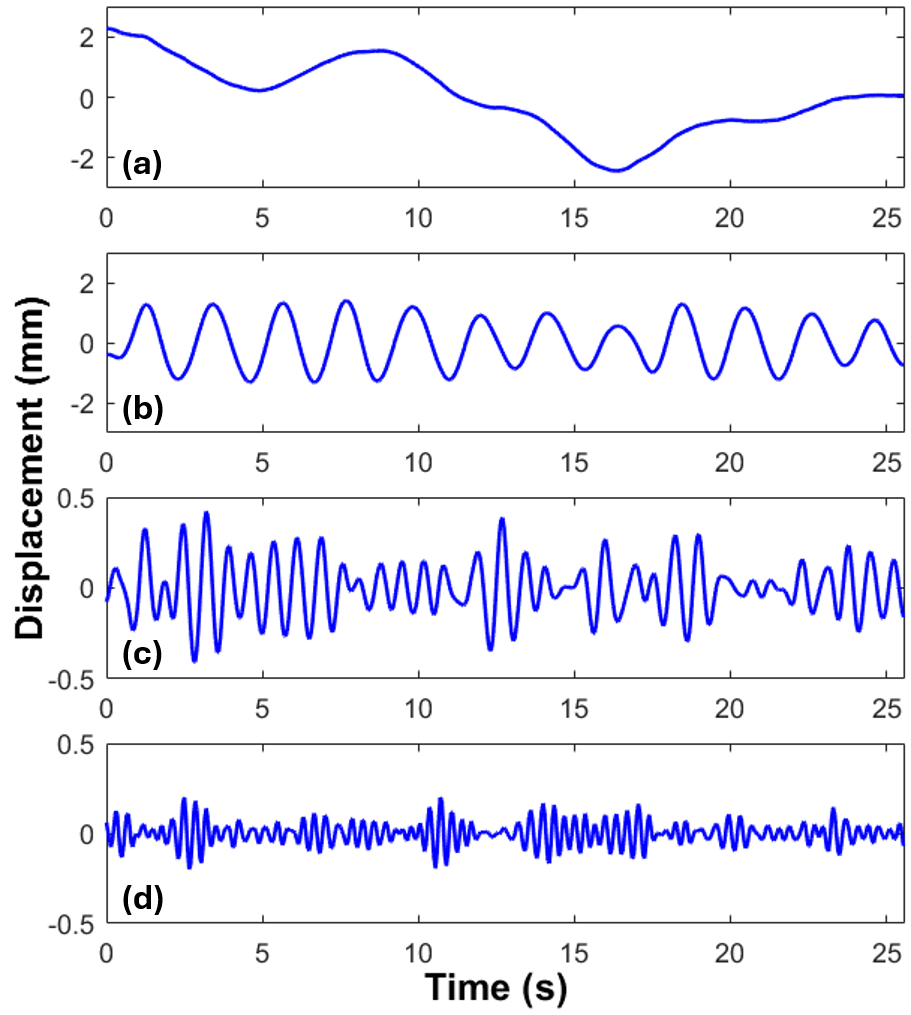}
        \caption{VMD Output - \textit{(a)} Low Frequency Noise \textit{(b)} Extracted raw BR signal \textit{(c)} Extracted raw HR signal \textit{(b)} High Frequency Noise}
    \label{VMD_result}
\end{figure}

\begin{figure}[]
    \centering
    \includegraphics[width=0.85\linewidth]{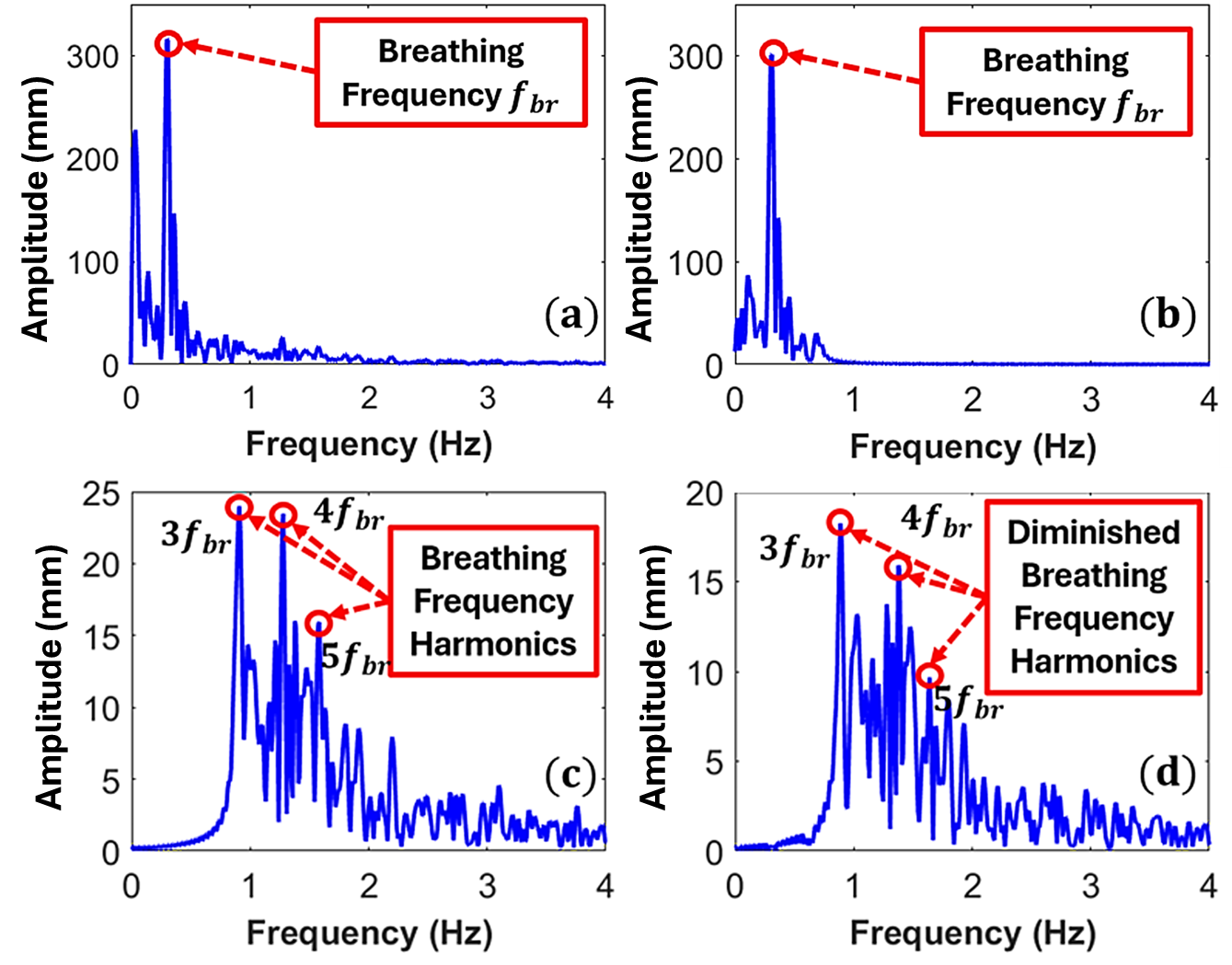}
    \caption{Frequency spectrum of \textit{(a)} unwrapped phase signal \textit{(b)} breathing signal \textit{(c)} heartbeat signal \textit{(d)} filtered heartbeat signal, received over a 25.6 s observation period}
    \label{extracted_signals}
\end{figure}

The typical BR of a human is in the range 3-36 breaths per minute (BRPM) and the typical HR is in the range 48-120 beats per minute (BPM), respectively \cite{fmcw_radar1}. The chest displacement signal $x(t)$ can be expressed as shown in Eq. \ref{eq:x(t)}, where $x_{br}(t)$ and $x_{hr}(t)$ are the breathing and heartbeat chest displacement signals, and $\eta (t)$ is the noise term containing a low-frequency band noise term $\eta _l(t)$ and a high-frequency band noise term $\eta_h(t)$.
\begin{equation} \label{eq:x(t)}
    x(t) = x_{br}(t) + x_{hr}(t) + \eta (t)
\end{equation}
During a brief observation window, breathing and heartbeat signals exhibit predominantly deterministic behavior \cite{deterministic}. Utilizing this, the unwrapped phase signal can be decomposed into its components, i.e., breathing signal, heartbeat signal, and noise using VMD.
VMD splits a real signal into a discrete number of sub-signals or modes ${\mu_k}$ with center frequencies ${\omega_k}$, where $k = 1, 2, ..., K$.
The solution to the problem is the optimization of Eq. \ref{eq:VMD} \cite{vmd}. With the assumption that $x(t)$ purely contains four modes, the VMD solution can be optimized for $\{\mu_k(t) \}$ = $\{x_{br}(t), x_{hr}(t), \eta_l(t), \eta_h(t)\}$.
\begin{equation} \label{eq:VMD}
\begin{split}
    \mathcal{L}(\mu_k, \omega_k, \lambda) = \alpha \sum \limits_{k=1}^{K} \Big|\Big| \frac{d}{dt}\Big[\Big(\delta(t) + \frac{j}{\pi t}\Big) * \mu_k(t)\Big]e^{-j\omega_kt} \Big|\Big|_2^2 \\ + \Big|\Big|x(t) - \sum \limits_{k=1}^{K} \mu_k(t) \Big|\Big|_2^2 + \Bigg<\lambda(t),x(t) - \sum \limits_{k=1}^{K} \mu_k(t) \Bigg>
\end{split}
\end{equation}
\textcolor{Black}{Fig. \ref{VMD_result} shows the VMD of a sample unwrapped phase signal into its four modes}.
Breathing harmonics can overlap with true heartbeat frequencies, overshadowing them due to the significantly lower magnitude of the heartbeat signal. Appropriate comb filters with notches at the breathing harmonics can recover the true heartbeat signal. Fig. \ref{extracted_signals} illustrates the frequency spectra of the extracted signal, 
showing pronounced peaks
at the breathing frequency and its harmonics, which appear as peaks in the heartbeat signals.

While extracting the breathing frequency is straightforward, extracting the heartbeat frequency is challenging. The noise term $\eta(t)$, which includes environmental noise, electronic noise, and motion artifacts, has a power similar to $x_{hr}(t)$ and can harshly affect the heartbeat frequency spectrum even after decomposing the unwrapped phase signal.
We propose the following three methods to extract the HR/BR features.

\subsubsection{Welch Spectrum} \label{subsec:Welch}
The Welch method for frequency estimation provides improved frequency resolution, reduced spectral leakage, and robustness to noise compared to traditional spectrum estimation techniques \cite{oppenheim, hayes}. While the breathing signal predominantly has one major peak, the heartbeat signal may feature multiple peaks. Thus, the highest peak for the breathing signal and a weighted average of $l$ highest peaks for the heartbeat signal and comb filtered heartbeat signal are taken as the Welch frequency estimates - $br_w$, $hr_w$, and $hr_{wc}$\textcolor{Black}{, respectively}.

\subsubsection{Auto-correlation}  \label{subsec:Auto-correlation}
Auto-correlation identifies consistent temporal patterns and reduces transient noise, effectively estimating the time period of the original signal. By taking the inverse of this period, we obtain auto-correlation frequency estimates - $br_a$ for the breathing signal, $hr_a$ for the heartbeat signal, and $hr_{ac}$ for the comb filtered heartbeat signal.

\subsubsection{Peak Detection}  \label{subsec:Peak Detection}
Respiration and heartbeat cycles are characterized by peaks in their respective signals, mitigating low-frequency noise. The number of peaks in an observation window, divided by the total observation time, provides the peak detection frequency estimates $br_p$, $hr_p$, and $hr_{pc}$ for the breathing signal, heartbeat signal, and comb filtered heartbeat signal, respectively.

\subsection{Regression Model Training for HR/BR Estimation} \label{subsec:Regression Models}

The three breathing frequency estimates $br_w$, $br_a$ and $br_p$, and the six heartbeat frequency estimates $hr_w$, $hr_a$, $hr_p$, $hr_{wc}$, $hr_{ac}$, $hr_{pc}$ can be used as features to train a regression model against ground truth breathing rates and heart rates.
\textcolor{Black}{Given the limitations in interpreting other metrics \cite{R2_best} and the common use of $R^2$ score (coefficient of determination) and $MAE$ in regression analysis for physiological measurements using radars \cite{r2_mae_1, r2_mae_2, r2_mae_3, r2_mae_4}, we selected these two metrics for evaluation, given by Eq. \ref{eq:R^2} and Eq. \ref{eq:MAE}\textcolor{Black}{, respectively},}
\begin{equation} \label{eq:R^2}
    R^2 = 1 - \Big(\sum \limits_i (z_i - \hat{z_i})^2\big/\sum \limits_i (z_i - \Bar{z_i})^2\Big)
\end{equation}
\begin{equation} \label{eq:MAE}
MAE = \frac{1}{n} \sum_{i=1}^{n} |z_i - \hat{z}_i|
\end{equation}
where $z_i$ is the actual label, $\hat{z_i}$ is the predicted label and $\Bar{z_i}$ is the mean of the actual labels. $R^2$ score measures the linearity of the feature variables against the labels. A higher $R^2$ score indicates that the features can more effectively explain or predict the labels than a model with a lower $R^2$ score. $MAE$ can measure how well the model predicts the target variable. 
A lower MAE indicates better model performance.

\section{System Configuration and Constraints}
\label{sec:Practical Constraints}

This section discusses the system setup for estimating HR and BR of a subject, along with the radar configuration. The radar configuration imposes inherent hardware limitations on subject detection, which are discussed next. 
After localizing the subjects, estimating HR and BR faces some empirical limitations, as discussed in the consequent section.


\subsection{System Setup and Configuration}

\begin{figure}[]
    \centering
    \includegraphics[width=0.8\linewidth]{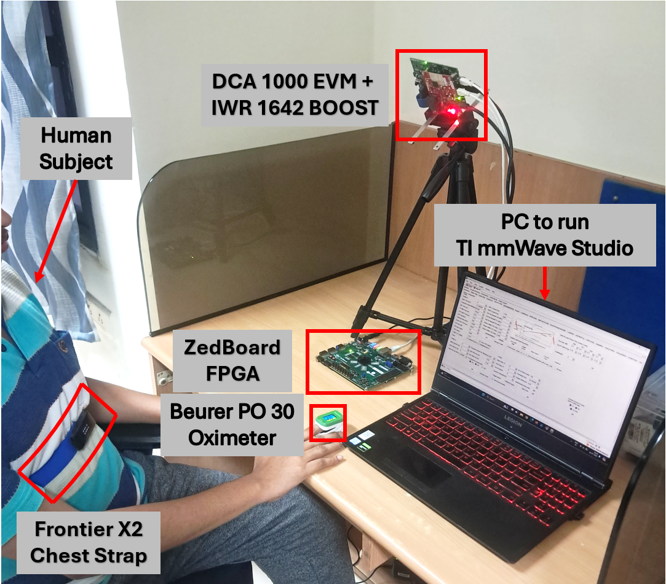}
    \caption{System Setup for measuring HR and BR}
    \label{fig:Experimental_Setup}
\end{figure}

Fig. \ref{fig:Experimental_Setup} shows the overall system setup to measure HR and BR of a subject, which includes the FMCW Radar IWR1642BOOST \cite{TI_IWR1642BOOST} and data capture evaluation module DCA1000EVM \cite{TI_DCA} from Texas Instruments (TI). TI's mmWave Studio \cite{TI_MMWAVE-STUDIO} software facilitates communication between radar and a PC, allows custom configuration of system and chirp parameters, and collects data for processing. Beurer PO 30 Oximeter \cite{BeurerPO30} and Frontier X2 Chest Strap \cite{FrontierX2} are used to measure the subjects' ground truth HR and BR. 
Additionally, the subjects manually counted their breaths throughout the experiments for verification. 

The proposed radar data processing pipeline is initially developed using MATLAB R2024a and Python v3.12.0. 
It is then successfully implemented on the ZedBoard FPGA \cite{avnet_zedboard} hardware platform, serving as a proof of concept for a portable and seamless integration solution. 
Table \ref{tab:system_params} shows our radar system configuration. 

Based on this configuration, a slow time sampling frequency ($\frac{1}{T_c}$) of 20 Hz allows for measuring chest displacement frequencies 
up to 10 Hz, well above typical human BR. The chosen number of chirps, M, sets the frequency resolution ($\frac{1}{MT_c}$) at 0.039 Hz, or 2.3 breaths per minute, an acceptable minimum for BR.


\begin{table}
\centering
\caption{Radar Module Configuration}
\label{tab:system_params}
{
\begin{tabular}{c c}
\hline 
\textbf{Radar Parameter} & \textbf{Value} \\ \hline \rowcolor{Gray}
Transmitters             & 2              \\ 
Receivers                & 4              \\  \rowcolor{Gray}
Start Frequency, $f_{min}$       & 77 GHz         \\ 
Stop Frequency, $f_{max}$           & 79.982 GHz    \\   \rowcolor{Gray}  
RF Bandwidth, $BW$             & 2998.2 MHz      \\    
Chirp Duration, $T_m$          & 100 $\mu$s \\    \rowcolor{Gray}
Chirp Period, $T_c$         & 50 ms \\ 
Chirp Slope, $K_c$              & 29.982 MHz/$\mu$s  \\ \rowcolor{Gray}
Number of Chirps, $M$        & 512 \\   
ADC Samples, $N$             & 512 \\ \rowcolor{Gray}
ADC Sampling Frequency, $f_{adc}$  & 6000 ksps
\\ \hline  
\end{tabular}}
\end{table}


\subsection{Hardware Limitations on Subject Detection}
Given a specific radar configuration, several inherent limitations affecting subject detection come into effect, as outlined in the following subsections.


\subsubsection{Maximum Detectable Range}
Maximum detectable range $R_{max}$ is influenced by the radar's bandwidth $BW$, chirp duration $T_m$, and sampling frequency $f_{adc}$ and can be calculated using Eq. \ref{eq:Rmax}. However, this range is constrained by frequencies aliased beyond half of the ADC sampling frequency, as per the Nyquist-Shannon sampling theorem \cite{Shannon}. Thus, the constrained maximum detectable range $R_{max}$ is given by Eq. \ref{eq:Rmax'}. 
\begin{align}
    R_{max}  & = c\frac{T_m}{2} \label{eq:Rmax}\\
    R_{max}' & = R_{max}\frac{f_{adc}}{2BW} \label{eq:Rmax'}
\end{align}
For the radar's configuration in Table \ref{tab:system_params}, $R_{max}$ is calculated to be 15 m, beyond which a subject will not be detected.

\subsubsection{Range Resolution Limit}
Equation \ref{eq:Rmin} gives the range resolution $R_{min}'$, which is limited by the number of ADC samples $N$.
\begin{align}
    R_{min}' = \frac{R'_{max}}{N/2} \label{eq:Rmin}
\end{align}
For the radar's configuration given in Table \ref{tab:system_params}, $R_{min}'$ is calculated to be 0.058 m.
This is well within the accepted limit as the depth of a human body is much greater than the range resolution, ensuring that two subjects cannot be closer to each other than the range resolution. 
It also ensures that no static reflectors are present in the same bin, with static body parts being the sole potential source of DC offset contribution, which is negligible.


\subsubsection{Angle Resolution Limit} \label{angle_res_limit}

The radar’s angle resolution depends on the number of TX-RX pairs used. Increasing this resolution distinguishes IF signals in the same range and nearby azimuth bins. In Single-Input-Multiple-Output (SIMO) radar, improving angle resolution often means doubling the receive antennas. \textcolor{Black}{Multiple-Input-Multiple-Output} (MIMO) radar and techniques like Time-Division Multiplexing offer ways to further enhance resolution \cite{fmcw_radar2}. 
Our radar is a MIMO radar, and its configuration, as shown in Table \ref{tab:system_params}, provides a 15$\degree$ angle resolution. Subjects with smaller angular separations are treated as a single entity.
This limits subject density in crowded environments, crucial for accurately detecting subjects.

\subsubsection{Minimum Elevation for Line-of-Sight Measurement in Multi Subject Scenario} \label{subsec:LoS Meas.}
\begin{figure}
    \centering
    \includegraphics[width=0.7\linewidth]{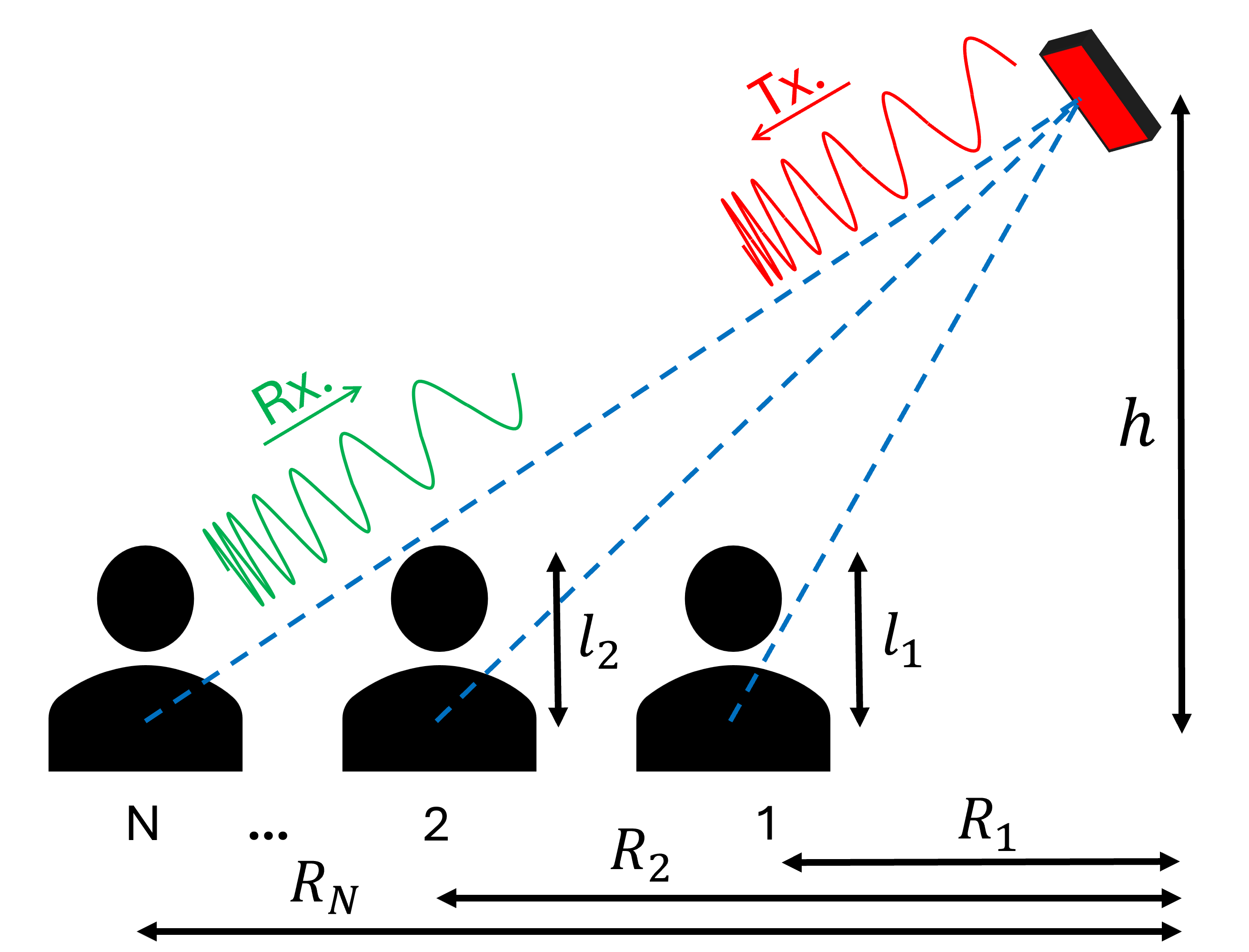}
    \caption{Proposed setup with radar at an elevated height}
    \label{fig:radar_at_height}
\end{figure}

Millimeter waves have restricted penetration through human body \cite{FMCW_penetration_power}. 
When radar is positioned at the chest level, as seen in prior works \cite{fmcw_radar2, fmcw_radar3, fmcw_radar4, fmcw_radar5, prev_work}, it is challenging to detect signals from subjects obscured by the first subject in the azimuth bin.
To overcome this, we propose to elevate the radar and tilt it towards the subjects as shown in Fig. \ref{fig:radar_at_height}.
The minimum elevation $h_i$ required for unobstructed line of sight to the $i^\text{th}$ subject can be calculated using Eq. \ref{eq:h_i}. 
\begin{equation} \label{eq:h_i}
    h_i = 
    \begin{cases}
    \max\big(l_{j}\frac{R_{i}}{R_i-R_{j}} \,|\, i > j \geq 1\big) & i > 1 \\
    0  & i = 1
    \end{cases}
\end{equation}
Since this must hold for all $i = 1, ...N$, the minimum elevation is given by $h_{min} = \max(h_1, h_2, ...h_N)$. 
This allows the detection of obscured subjects but imposes empirical limitations on HR and BR estimation, discussed in the following subsection.


\subsection{Empirical Limits of the proposed HR and BR Estimation Algorithm}

In the previous subsection, we discussed hardware limitations specifically affecting subject detection. 
Post-subject detection, practical HR and BR estimation are impacted by environmental noise, multi-path reflection, and increased SNR \cite{radar_eq_ref}. 
We empirically establish algorithm thresholds for precise HR and BR estimation, defining it as \textit{BR MAE $>$ 2}.
Details of these experiments are outlined below.

\subsubsection{Subject Distance Variation}
\label{radar_range}



\begin{figure}
    \centering
    \includegraphics[width=0.83\linewidth]{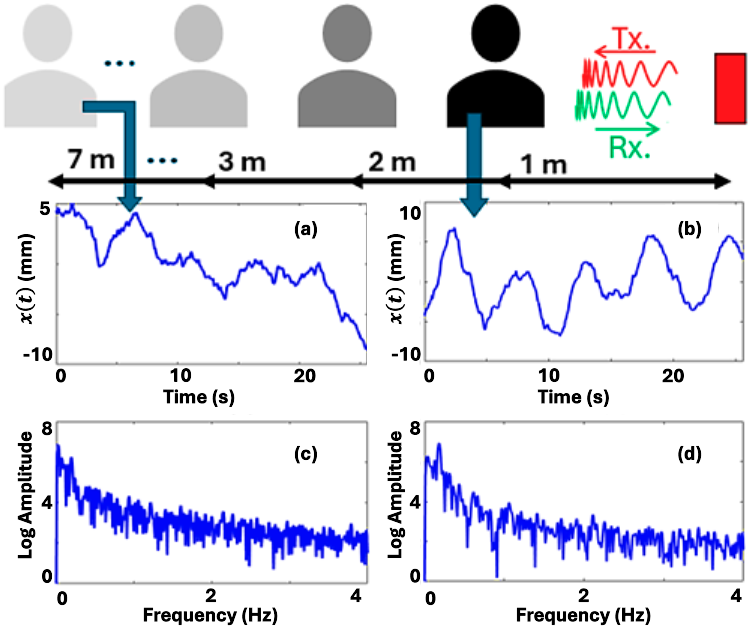}
    \caption{Impact of subject distance variation on \textit{(a)-(b)} phase signal and \textit{(c)-(d)} frequency spectrum}
    \label{fig:vs_range}
\end{figure}



Fig. \ref{fig:vs_range} shows the experimental setup to examine how subject distance variation affects HR and BR estimation,
with the subject seated at distances from 1 to 7 m.
Analysis of the phase signal plots and their frequency spectra 
shown in Fig. \ref{fig:vs_range}\textit{(a)-(d)}
shows increasing distortion with distance. The breathing peak becomes harder to distinguish amidst low-frequency noise, especially at 7 m with \textit{BR MAE $=$ 2.34}, as shown in Table \ref{table:expt_results}.
Notably, high-frequency noise, particularly in the heartbeat range, introduces unnecessary peaks in the phase signal, 
impacting the accuracy of HR estimation. 



\subsubsection{Subject Azimuth Variation}
\begin{figure}
    \centering
    \includegraphics[width=1\linewidth]{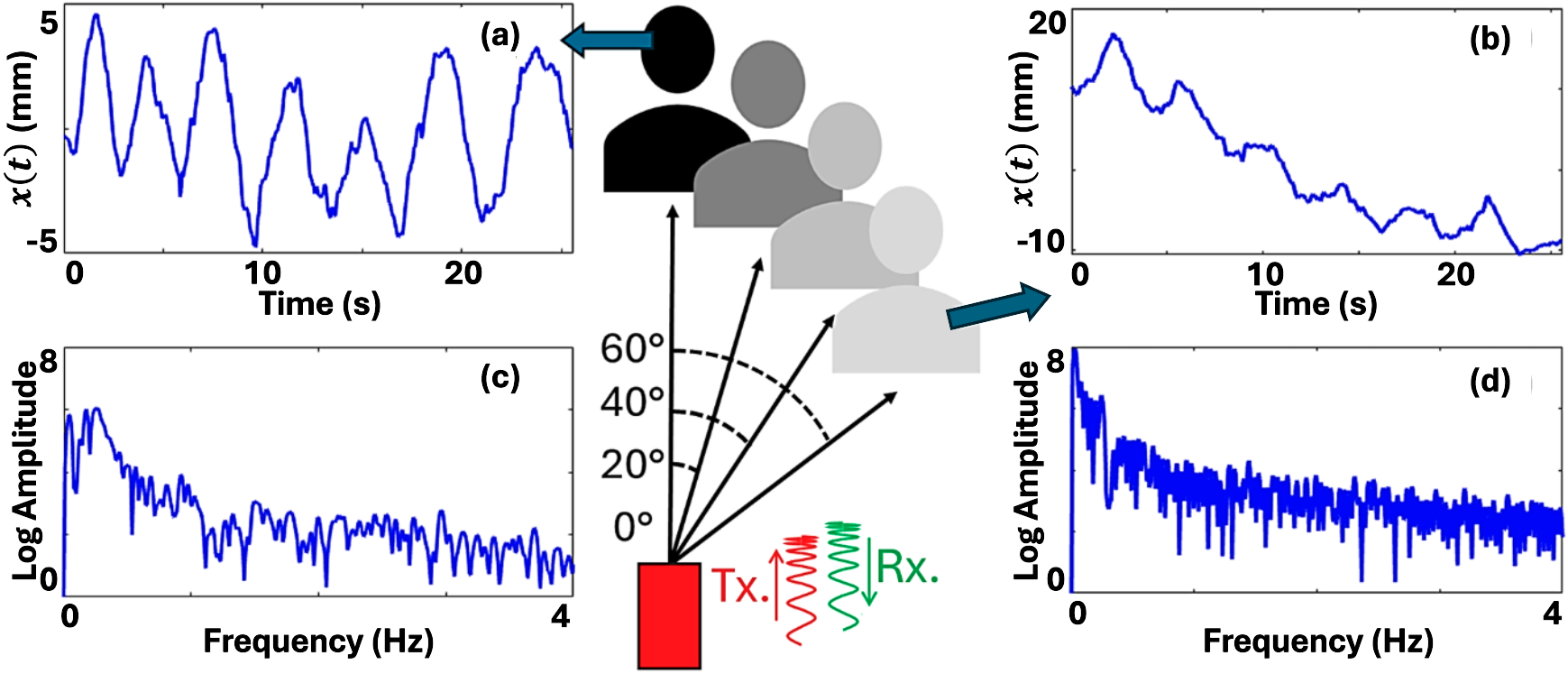}
        \caption{Impact of subject azimuth variation on \textit{(a)-(b)} phase signal and \textit{(c)-(d)} frequency spectrum}
    \label{fig:vs_azi}
\end{figure}

Fig. \ref{fig:vs_azi} shows the experimental scenario to study subject azimuth variation's effect on HR/BR estimation. The subject was seated at a fixed distance of 1 m, but at azimuth increments of $\pm$20$\degree$ from 0$\degree$ to $\pm$60$\degree$. It is observed 
from Fig. \ref{fig:vs_azi}\textit{(a)-(d)}
that the cleanest phase signal is captured when the subject was at 0$\degree$.
The signal spectrum becomes noisier as the subject azimuth increases. 
At
higher azimuths, such as $\pm$60$\degree$, the baseline wander and mid-frequency noise in the signal becomes very evident,
significantly affecting phase signal quality and resulting in \textit{BR MAE $=$ 3.39}, as shown in Table \ref{table:expt_results}.

\subsubsection{Subject Posture Variation} \label{posture_variation}
\begin{figure}
    \centering
    \includegraphics[width=0.83\linewidth]{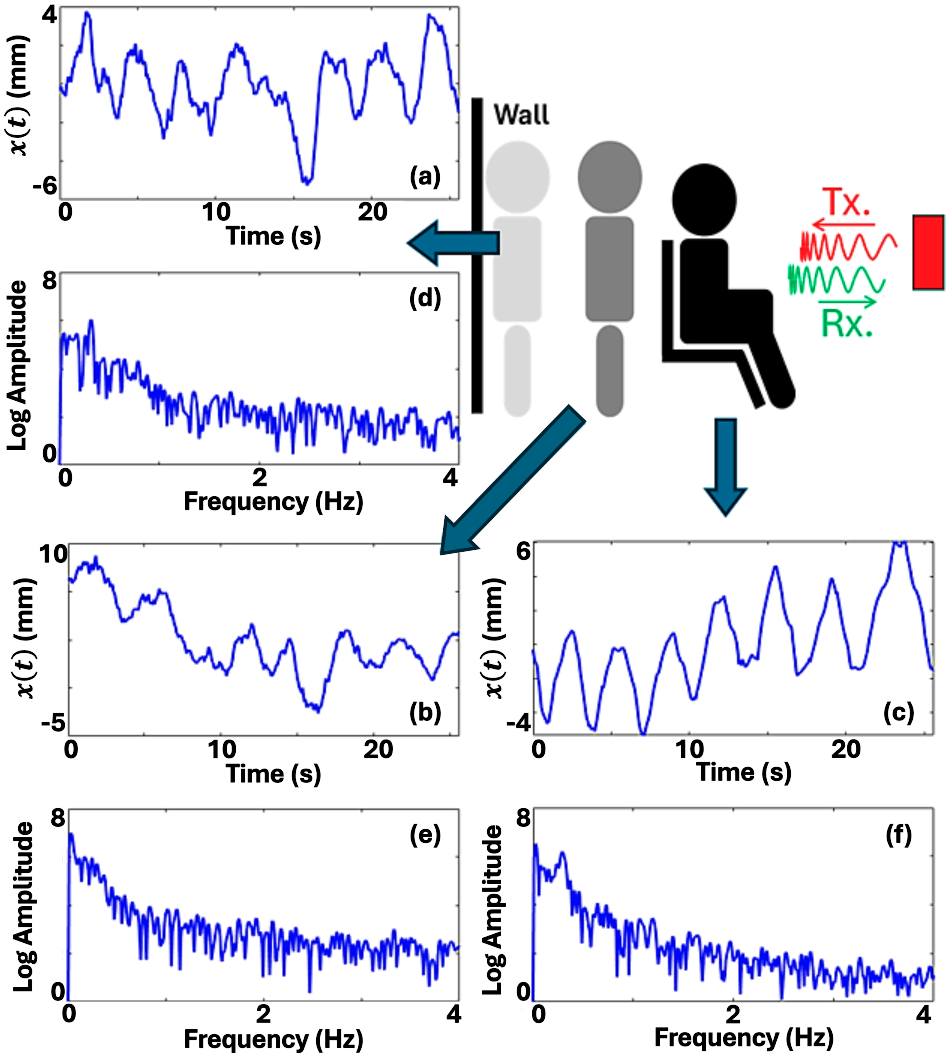}
        \caption{Impact of subject posture variation on \textit{(a)-(c)} phase signal and \textit{(d)-(f)} frequency spectrum}
    \label{fig:vs_pos}
\end{figure}

As shown in Fig. \ref{fig:vs_pos}, subjects are made to sit, stand, and stand against a wall, across experiments, at a consistent distance of 1 m from the radar. 
As seen in Fig. \ref{fig:vs_pos}\textit{(a)-(f)}
the cleanest phase signals are obtained when the subject is seated. 
In contrast, standing without support causes slight body movements that distort the phase signal significantly, resulting in the worst \textit{BR MAE $=$ 4.44} as shown in Table \ref{table:expt_results}. 
However, standing with support increases stability, yielding cleaner phase signals, though not as clean as sitting.



\subsubsection{Radar Elevation Variation}
\label{radar_elev}

\begin{figure}
    \centering
    \includegraphics[width=1\linewidth]{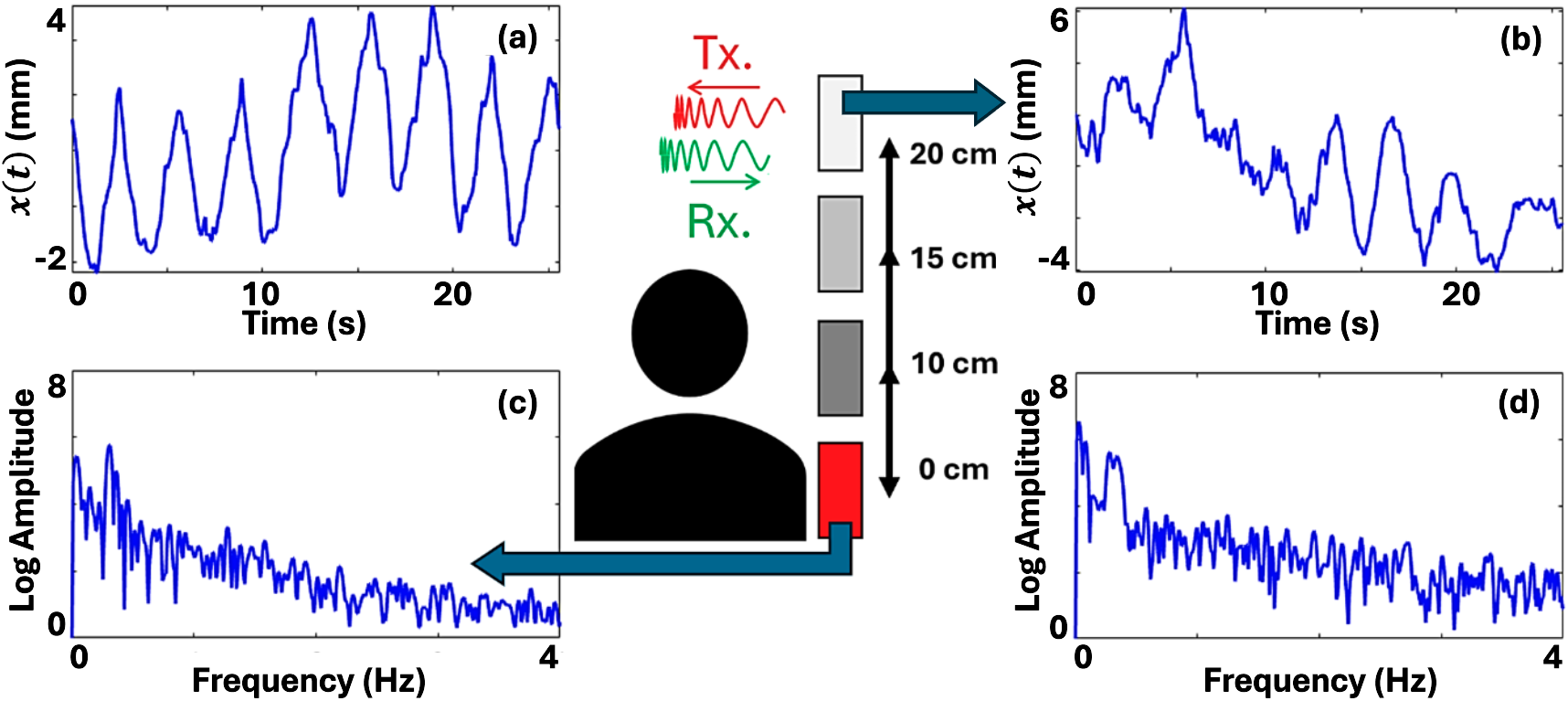}
        \caption{Impact of radar elevation variation on \textit{(a)-(b)} phase signal and \textit{(c)-(d)} frequency spectrum}
    \label{fig:vs_elev}
\end{figure}

As discussed in Subsection \ref{subsec:LoS Meas.}, an elevated radar system is crucial for unobstructed HR and BR measurement of multiple subjects. Fig. \ref{fig:vs_elev} illustrates the experiment to examine the trade-off between radar elevation and HR and BR measurement accuracy. The subject is seated 1 m away from the radar. 
As the radar's elevation increases, the line-of-sight distance from the subject increases and the angle with the chest wall becomes steeper, both leading to a significant decrease in $\text{SNR}$ and phase signal quality.
As seen in Fig. \ref{fig:vs_elev}\textit{(a)-(d)},
this effect is particularly noticeable when comparing phase signal quality between radar positions at chest level and 20 cm above chest level. 
At 20 cm elevation, the phase signal shows significant white noise, due to the radar picking up additional environmental interference not focused on the chest wall, resulting in \textit{BR MAE $=$ 3.65}, as shown in Table \ref{table:expt_results}.
\begin{figure}[!h]
    \centering
    \includegraphics[width=1\linewidth]{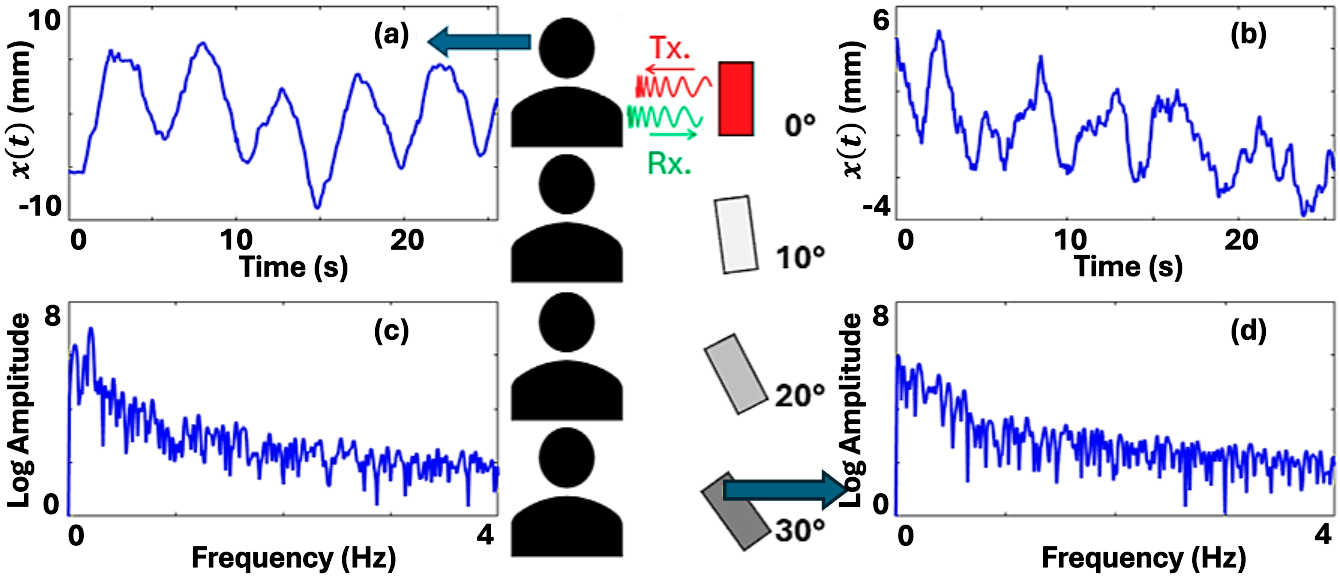}
        \caption{Impact of radar tilt variation on \textit{(a)-(b)} phase signal and \textit{(c)-(d)} frequency spectrum}
    \label{fig:vs_tilt}
\end{figure}

\subsubsection{Radar Tilt Variation}
\label{radar_tilt}

Fig. \ref{fig:vs_tilt} shows the experimental scenario to study the trade-off between radar tilt and HR/BR measurement accuracy. 
The subject is seated at a fixed distance of 1 m, with radar tilt ranging from 0$\degree$ to 30$\degree$ in 10$\degree$ increments.
As the radar tilt increases, the angle between the radar and the chest wall of the subject increases. Similar to Subsection \ref{radar_elev}, as the radar faces further away from the chest wall, the effect of environmental white noise on the extracted phase signal increases. 
As seen in Fig. \ref{fig:vs_tilt}\textit{(a)-(d)},
the cleanest phase signal is obtained when the radar tilt is 0$\degree$, i.e., directly facing the chest wall of the subject. 
At 30$\degree$ tilt, white noise is significant and this drastically affects the SNR of the phase signal, resulting in \textit{BR MAE $=$ 2.52}, as shown in Table \ref{table:expt_results}.

Table \ref{table:expt_results} summarizes the results of the above-mentioned experiments.
For each \textcolor{Black}{ experimental variable, the results show an increasing trend in the MAE measured for the Welch estimate 
of
BR and HR, and we have concluded the experiments in each case at the point where \textit{BR MAE $>$ 2}}.

\begin{table}
\centering
\caption{MAE of BR/HR Welch Estimate Against Experiment Variables}
\label{table:expt_results}
\resizebox{\columnwidth}{!}{
\begin{tabular}{cccc}
\hline
\textbf{\begin{tabular}[c]{@{}c@{}}Expt \\ Variable\end{tabular}} & \textbf{Value} & \textbf{\begin{tabular}[c]{@{}c@{}}BR MAE\\ (BRPM)\end{tabular}} & \textbf{\begin{tabular}[c]{@{}c@{}}HR MAE\\ (BPM)\end{tabular}} \\
\hline \rowcolor{Gray}
 & 1 & 0.51 & 6.80 \\ \rowcolor{Gray}
 & 2 & 0.47 & 7.51 \\ \rowcolor{Gray}
 Subject  & 3 & 0.87 & 8.40 \\ \rowcolor{Gray}
 Distance  & 4 & 1.14 & 14.34 \\ \rowcolor{Gray}
 (m) & 5 & 0.94 & 13.41 \\ \rowcolor{Gray}
 & 6 & 1.31 & 16.53 \\ \rowcolor{Gray}
 & 7 & 2.34 & 17.89 \\
 \hline
 & 0 & 0.78 & 5.36 \\
 Subject Azimuth & $\pm$20 & 1.28 & 7.96 \\
 ($\degree$) & $\pm$40 & 1.24 & 12.13 \\
 & $\pm$60 & 3.39 & 11.90 \\
 \hline \rowcolor{Gray}
 & Sitting & 0.68 & 6.78 \\ \rowcolor{Gray}
 Subject Posture & Standing against wall & 1.45 & 10.81 \\ \rowcolor{Gray}
 & Standing & 4.44 & 15.33 \\
 \hline
 & 0 & 0.53 & 5.33 \\
 Radar Elevation& 10 & 0.93 & 7.98 \\
 (cm) & 15 & 1.75 & 11.39 \\
 & 20 & 3.65 & 13.77 \\
 \hline \rowcolor{Gray}
 & 0 & 0.71 & 6.33 \\ \rowcolor{Gray}
 Radar Tilt& 10 & 0.89 & 10.47 \\ \rowcolor{Gray}
 ($\degree$) & 20 & 1.75 & 10.43 \\ \rowcolor{Gray}
 & 30 & 2.52 & 14.67 \\
 \hline
\end{tabular}}
\end{table}

\section{System Implementation and Proposed Algorithm Performance} \label{sec:HW Implementation}


In this section, we begin by setting up a practical experimental setup, considering the constraints and trade-offs discussed in Section \ref{sec:Practical Constraints}, to study the algorithm's performance. We then discuss the software implementation followed by proof-of-concept hardware implementation on ZedBoard FPGA and evaluate our algorithm's performance against prior works.

\subsection{Experimental Setup for Evaluating the Proposed Algorithms}

\begin{figure}
    \centering
    \includegraphics[width=1\linewidth]{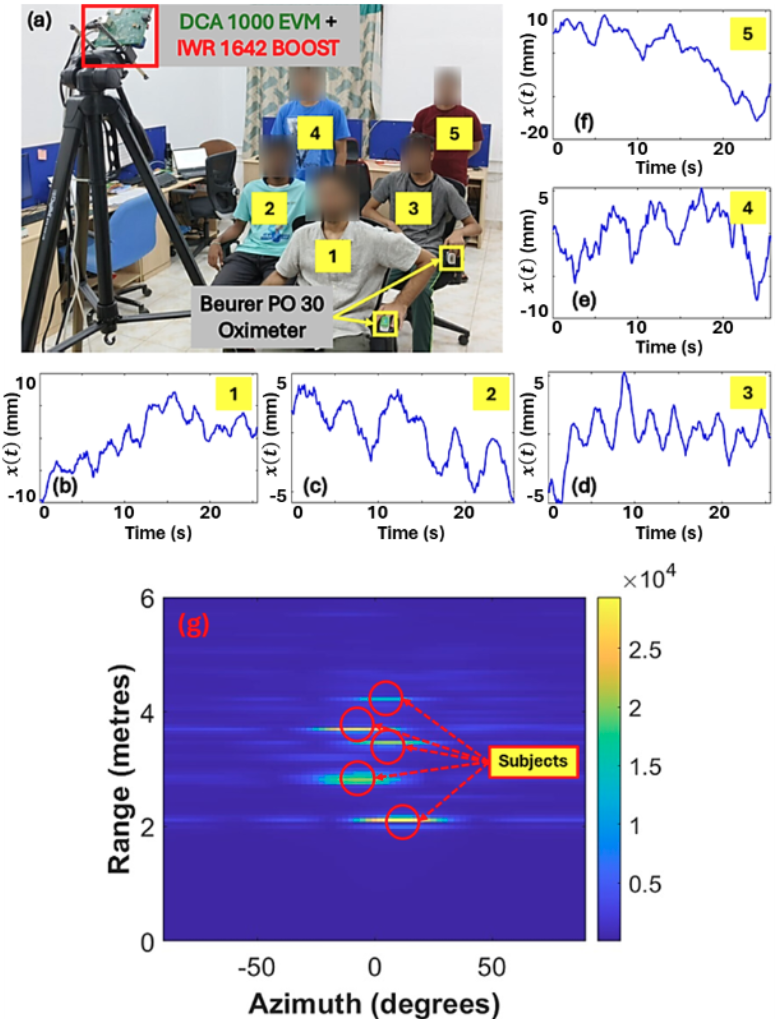}
        \caption{\textit{(a)} Experimental setup \textit{(b)-(f)} phase unwrapped signals of corresponding subjects and \textit{(g)} Range-Azimuth map}
    \label{fig:sit_stand_pus}
\end{figure}

Fig. \ref{fig:sit_stand_pus}(a) illustrates the experimental setup used to evaluate the proposed algorithm in realistic scenarios. To mimic real-world situations, the following considerations were made:

\begin{enumerate}
    \item \textbf{Range}: \textcolor{Black}{The empirical range of the algorithm limits the positioning of subjects to a maximum distance of 6 m, as discussed in Subsection \ref{radar_range}.}
    \item \textcolor{Black}{\textbf{Separation between subjects}: The hardware limits the separation between subjects to a minimum of 15°, as discussed in Subsection \ref{angle_res_limit}.}
    \item \textbf{Elevation for Line-of-Sight}: \textcolor{Black}{To address the limitation of subjects being obscured by others, the radar is elevated to a height of 2 meters, as discussed in Subsection \ref{subsec:LoS Meas.}}. 
    \item \textbf{Radar Tilt}: With the radar elevated, we strategically tilt it towards the center of the subjects' arrangement to mitigate elevation effects discussed in Subsection \ref{radar_elev}, minimizing inaccuracies. 
    \item \textbf{Arrangement for High Density}: To mimic high subject density seen in practical scenarios, subjects are arranged in a zig-zag pattern.
    \item \textbf{Subject Posture}: Three subjects were seated, and two \textcolor{Black}{ stood} without wall support to reflect \textcolor{Black}{common postures in real-world scenarios}.

\end{enumerate}
Fig. \ref{fig:sit_stand_pus}(b)-(f) demonstrates that seated individuals produced cleaner signals, similar to observations in Subsection \ref{posture_variation}. Additionally, the farther the subjects were from the center, the more the radar tilt and distance affected signal quality, leading to more distorted signals. \textcolor{Black}{Fig. \ref{fig:sit_stand_pus}(g) displays the corresponding Range-Azimuth map, where five subjects arranged in a \textcolor{Black}{ zig-zag pattern,}  are clearly distinguishable.}

The following subsections show how our algorithm performed in these practical scenarios for estimating HR and BR of the subjects and provide a comparison with prior works.
\subsection{Software Implementation of the Proposed Algorithms}
\subsubsection{Implementation}
As per Sections \ref{subsec:IF Signal Sampling} to \ref{subsec:HR/BR Feature Extraction}, MATLAB is used to extract HR and BR features of subjects which are subsequently used to estimate the final HR and BR using regression models trained in Python.
Fig. \ref{fig:sit_stand_pus} shows the practical arrangement of subjects from which raw data samples were collected to facilitate the training of regression models.
A dataset comprising around 400 samples of HR and BR features was created by processing this data in MATLAB. The ground truth (BR, HR) label pairs had a mean and standard deviation of $(18.42 \pm 7.38, 91.98 \pm 9.88)$. Regression models such as linear regressor (LR), random forest regressor (RF), support vector regressor (SVR), and K-nearest neighbors regressor (KNN) were used to learn the data. To evaluate the performance of the models, we are using two metrics - $R^2$ score and MAE. These models were trained on 80\% of the dataset to maximize the $R^2$ score and minimize the MAE.

\subsubsection{Results} \label{subsec:MATLAB_results}

The regression model performance was tested on the remaining 20\% of the HR/BR features dataset, with results summarized in Table \ref{tab:model_performance}. The RF regressor showed the best performance across all metrics for both BR and HR. RF regressors use an ensemble of decision trees, averaging their predictions to produce a final prediction that is more accurate and resilient to noise and outliers compared to a single decision tree. All regressors performed very well for BR prediction, with an MAE of less than one breath, mainly because the breathing signal and its features are less affected by noise due to larger chest displacements. 
\textcolor{Black}{Table \ref{table:expt_results_rf} summarizes the performance of \textcolor{Black}{the} RF regressor in estimating HR and BR across all experiment variables. 
The MAE for both HR and BR estimation increases as 
each variable's value increases.S
}
\begin{table}
\centering
\caption{HR/BR Prediction Performance of Various Regression Models}
\label{tab:model_performance}
\begin{tabular}{c c c}
\hline
\textbf{Model} & \textbf{$R^2$ (BR, HR)} & \textbf{MAE (BR, HR)} \\ \hline \rowcolor{Gray}
Linear regressor & (0.95, 0.22) & (0.98, 9.13) \\ 
Random forest regressor & (0.98, 0.72) & (0.39, 3.64) \\  \rowcolor{Gray}
Support vector regressor & (0.96, 0.16) & (1.00, 9.54) \\ 
K-nearest neighbours regressor & (0.95, 0.44) & (0.83, 5.98) \\ \hline
\end{tabular}
\end{table}
\begin{table}[]
\centering
\caption{MAE of BR/HR Random Forest Regression Model Estimate Against Experiment Variables}
\label{table:expt_results_rf}
\resizebox{\columnwidth}{!}{
\begin{tabular}{cccc}
\hline
\textbf{\begin{tabular}[c]{@{}c@{}}Expt \\ Variable\end{tabular}} & \textbf{Value} & \textbf{\begin{tabular}[c]{@{}c@{}}BR MAE\\ (BRPM)\end{tabular}} & \textbf{\begin{tabular}[c]{@{}c@{}}HR MAE\\ (BPM)\end{tabular}} \\
\hline \rowcolor{Gray}
 & 1 & 0.22 & 2.11 \\ \rowcolor{Gray}
 & 2 & 0.27 & 2.19 \\ \rowcolor{Gray}
Subject & 3 & 0.33 & 3.24 \\ \rowcolor{Gray}
Distance & 4 & 0.48 & 5.38 \\ \rowcolor{Gray}
(m) & 5 & 0.69 & 8.75 \\ \rowcolor{Gray}
 & 6 & 0.85 & 10.39 \\ \rowcolor{Gray}
 & 7 & 1.43 & 12.26 \\ \hline
 & 0 & 0.31 & 2.72 \\
Subject Azimuth & $\pm$20 & 0.34 & 3.01 \\
($\degree$) & $\pm$40 & 0.54 & 4.97 \\
 & $\pm$60 & 0.89 & 8.14 \\ \hline  \rowcolor{Gray}
 & Sitting & 0.20 & 2.45 \\ \rowcolor{Gray}
Subject Posture & Standing against wall & 0.47 & 4.81 \\ \rowcolor{Gray}
 & Standing & 1.11 & 9.28 \\ \hline
 & 0 & 0.24 & 2.73 \\
Radar Elevation & 10 & 0.41 & 3.82 \\
(cm) & 15 & 0.41 & 6.57 \\
 & 20 & 0.69 & 10.08 \\ \hline \rowcolor{Gray}
 & 0 & 0.27 & 2.31 \\ \rowcolor{Gray}
Radar Tilt & 10 & 0.32 & 5.93 \\ \rowcolor{Gray}
($\degree$) & 20 & 0.85 & 9.72 \\ \rowcolor{Gray}
 & 30 & 1.47 & 9.57 \\
 \hline
\end{tabular}}
\end{table}

\subsection{FPGA Implementation of the Proposed Algorithms} \label{subsec:FPGA}

\subsubsection{Implementation} \label{subsub:FPGA_implementation}

\begin{figure}
    \centering
    \includegraphics[width=1\linewidth]{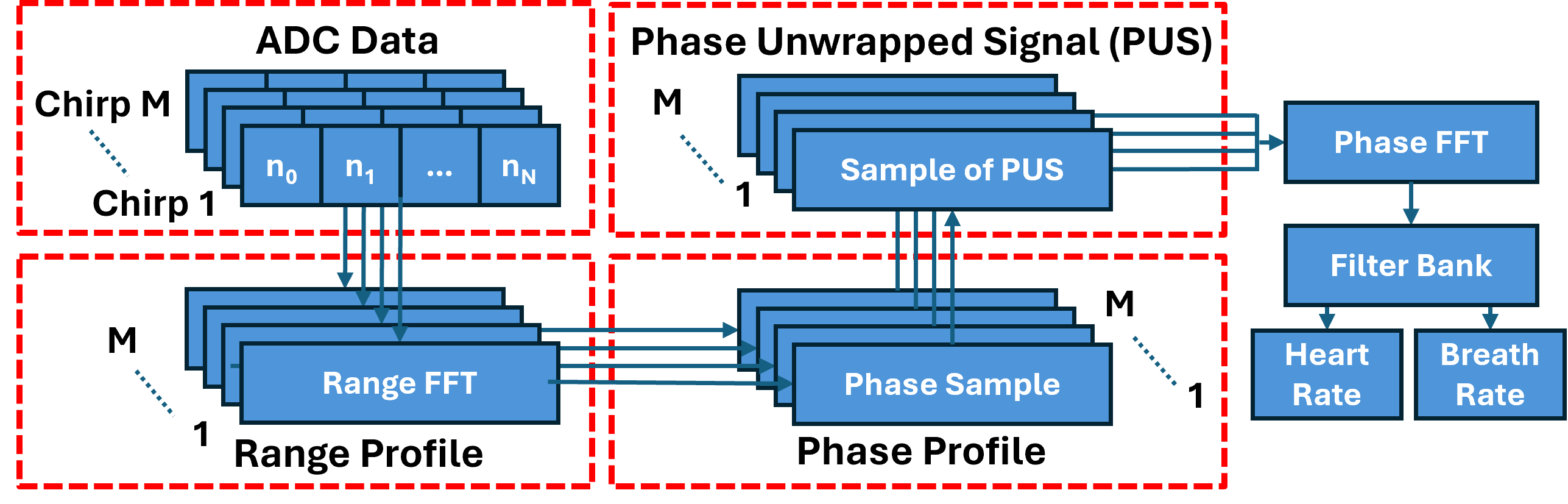}
    \caption{FPGA based signal processing flow}
    \label{fig:fpga flow}
\end{figure}

Fig. \ref{fig:fpga flow} outlines the design process from the acquisition of raw radar data to HR/BR measurement, with details of each step provided below.

\begin{enumerate}[label=(\alph*)]
    \item \textbf{Data Pre-processing} - Fig. \ref{fig:data_prepro_fpga} illustrates FPGA-based digitization of the IF signal from the DCA1000EVM board. We use 32-bit fixed-point representation to efficiently convert the analog signal to digital. Additionally, we pre-process the data to reduce sample size, addressing memory limitations and resource use. This processed data includes in-phase and quadrature-phase signals, and is given as input to Range FFT Module.
    \begin{figure}
        \centering
        \includegraphics[width=0.9\linewidth]{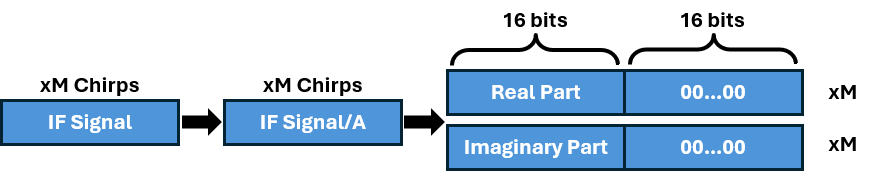}
        \caption{Radar Data Preprocessing Flow}
        \label{fig:data_prepro_fpga}
    \end{figure}
    \item \textbf{Range FFT Module} - Fig. \ref{range fft arch} shows a system-level overview of the RFFT module. The Range FFT (RFFT) module provides insights into the range or distance of subjects in the radar field and identifies the most appropriate index to target for further processing, by executing on the pre-processed digital data \cite{range_detection}.  
    The incoming IF signal from the radar is first stored in memory within the architecture. Then, the FFT operation is executed on this signal using the Xilinx FFT IP \cite{xilinxAdaptiveComputing} block within the FPGA design. This IP block is a pre-designed module that accelerates FFT computations, taking advantage of the FPGA's parallel processing capabilities \cite{xilinxAdaptiveComputing}. The power spectral density (PSD) of $X$ and Range of the subject is calculated using Eq. \ref{PSD_EQ} and \ref{eq_range}\textcolor{Black}{, respectively}.    
    \begin{equation} \label{PSD_EQ}
        \Gamma[k] = \left[Re(X^2[k]) + Im(X^2[k])\right]
    \end{equation}
    \begin{equation} \label{eq_range}
    R = \Big(\frac{1}{2} \times \frac{c}{BW} \times T_m\Big) \times f_{b} = K \times f_{b}
    \end{equation}
    The PSD aids in identifying significant frequency components, and the frequency index of the peak amplitude of $\Gamma$, denoted by $\hat{k}$, is computed to estimate the peak frequency $f_{b}$, with which the range of the object is estimated based on the known characteristics of the radar system.
    \begin{figure}
      \centering
        \includegraphics[width=0.9\linewidth]{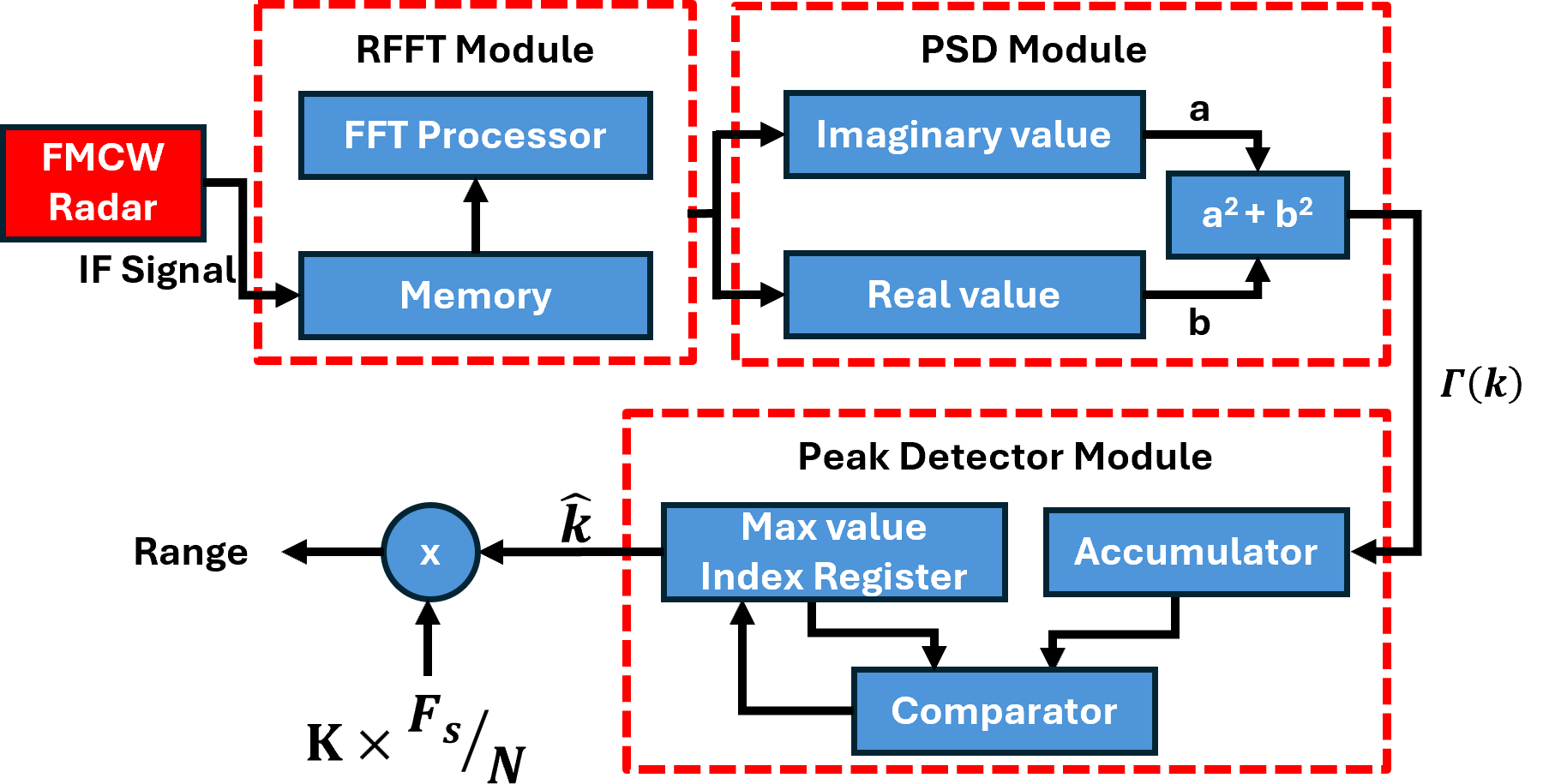}
      \caption{The architectural block diagram of the proposed RFFT module}
      \label{range fft arch}
    \end{figure}
    \item \textbf{Phase Extraction Module} - At the range computed by the RFFT module, the phase information is extracted from $X$, which contains chest displacement information. 
    The CORDIC (coordinate rotation digital computer) algorithm implemented via the Xilinx IP block \cite{xilinxAdaptiveComputing} is used to extract phase information by efficiently computing trigonometric functions. 
    The phase information is stored in a Block RAM (BRAM) to ensure high-speed and efficient storage, and is input to the Phrase Unwrapping Module.
    The method is customized for our system, involving specific configurations like setting the rotation angle, choosing appropriate data representation, and optimizing resource efficiency. 
    \item \textbf{Phase Unwrapping Module} - 
    Fig. \ref{unwrapping} shows the various components of unwrapping module architecture \cite{our_ises}.
    The phase information is fetched sample-by-sample from BRAM into an accumulator. 
    A dedicated register retains the previous phase value, serving as a reference for comparing consecutive phase samples. The differentiation circuitry computes the rate of change between these samples by taking inputs from the previous value register and the current phase sample. The comparator evaluates this difference, checking whether it falls within the range of $-\pi$ to $\pi$, failing which, adjustments are made to keep the phase values within the range of $-\pi$ to $\pi$. This adjusted value is saved to the output register and fed back into the accumulator for continuous comparison until the difference between consecutive samples is less than $\pi$.
    \begin{figure}
        \centering
        \includegraphics[width=1\linewidth]{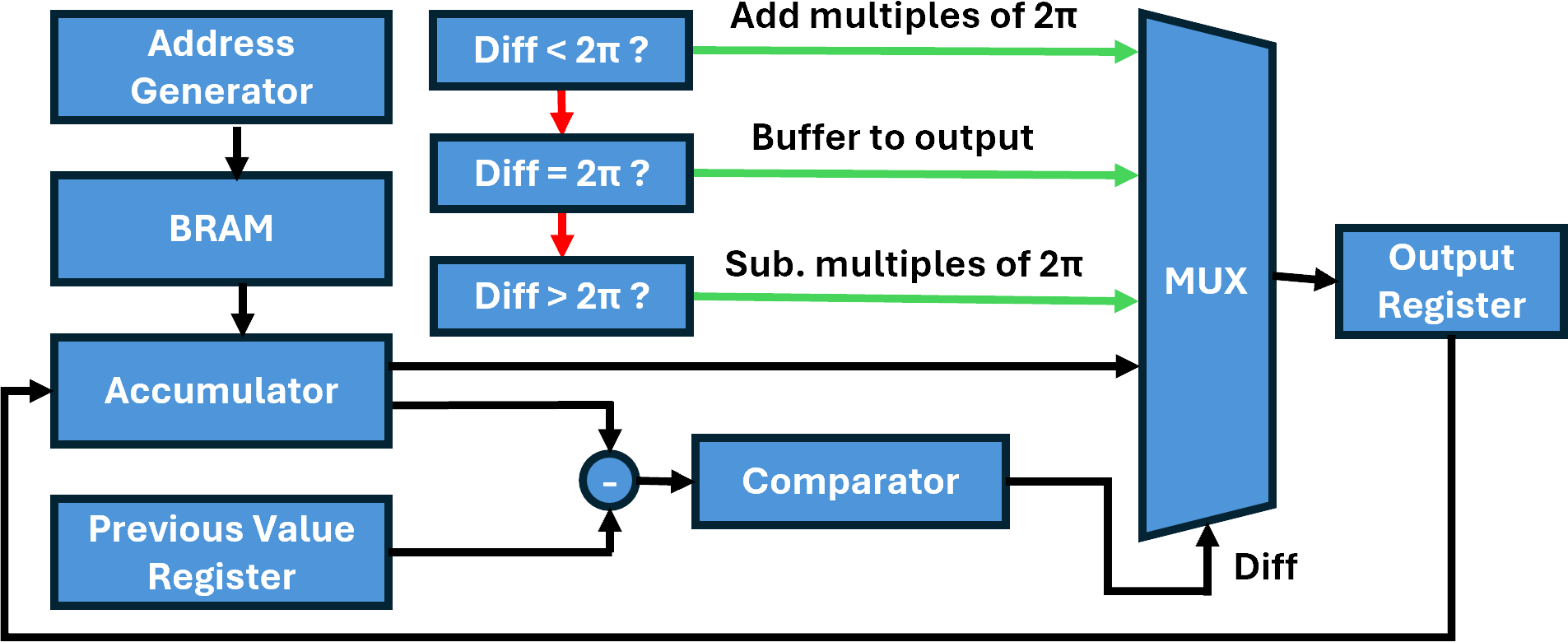}
        \caption{Phase Unwrapping Module Architecture}
        \label{unwrapping}
    \end{figure}
    \item \textbf{Phase FFT and HR/BR Estimation Module} - The output of the phase unwrapping module contains HR/BR information but may include noise and additional frequencies. The Phase FFT (pFFT) module computes the spectrum of the unwrapped phase signal. Next, HR/BR is estimated by analyzing peaks of the spectrum. To conserve hardware resources, 
    we skip FIR and IIR filters, opting to divide the FFT into bins and scan corresponding HR and BR bands.
    The normalized peak index in these bands multiplied by the sampling frequency gives the corresponding HR and BR. 
\end{enumerate}
\begin{table}[]
\centering
\caption{Results for HR/BR Estimation Using ZedBoard FPGA}
\label{table:results}
\resizebox{\columnwidth}{!}{
\begin{tabular}{ccccc}
\hline 
 & \textbf{Result 1} & \textbf{Result 2} & \textbf{Result 3} & \textbf{Result 4}\\ \hline \rowcolor{Gray}
\textbf{True HR (BPM)} & 87 & 105 & 104 & 77\\ 
\textbf{Estimated HR (BPM)} & 82 & 101 & 98 & 73 \\ \rowcolor{Gray}
\textbf{True BR (BRPM)} & 18 & 21 & 12 & 10 \\
\textbf{Estimated BR (BRPM)} & 14 & 23 & 14 & 13 \\ \hline
\end{tabular}}
\label{table: result table hr br}
\end{table}

\begin{table}[]
\centering
\caption{Resources and latency comparison of  FPGA based signal processing of radar signals.}
\label{fpga resouce comaprarion}
\resizebox{\columnwidth}{!}{
\begin{tabular}{ccccc}
\hline
& \multicolumn{1}{c}{\textbf{This work}} & \multicolumn{1}{c}{\textbf{\cite{tudose2018pulse}}} & \multicolumn{1}{c}{\textbf{\cite{tsao2018two}}} & \multicolumn{1}{c}{\textbf{\cite{heo2021fpga}}} \\ \hline
\rowcolor{Gray}
\multicolumn{1}{c}{\textbf{Type of radar}} & FMCW & Pulse & CMOS impulse & FMCW  \\ 
\multicolumn{1}{c}{\textbf{LUT}} & 8889 & 168,619 & 207,267 & 10891 \\ \rowcolor{Gray}
\multicolumn{1}{c}{\textbf{FF}} & 13805 & 60,547 & 6535 & 6365 \\ 
\multicolumn{1}{c}{\textbf{DSP}} & 24 & 1,540 & 560 & 20\\ \rowcolor{Gray}
\multicolumn{1}{c}{\textbf{Clock}} & 300 MHz & 320 MHz & 318 MHz & 300 MHz  \\ 
\textbf{Execution time} & 0.815ms & - & 35.4ms & 3ms\\ \hline
\end{tabular}}
\end{table}

\begin{table*} [!ht]
\centering
\caption{Comparison of HR/BR Measurement Performance}
\resizebox{\linewidth}{!}{
\begin{tabular}{ccccccccccccccccccc}
\hline
 & \multirow{2}{*}{\textbf{\begin{tabular}[c]{@{}c@{}}This Work \\(MATLAB\textcolor{Black}{+Python})\end{tabular}}} & \multirow{2}{*}{\textbf{\begin{tabular}[c]{@{}c@{}}This Work\\(FPGA)\end{tabular}}} & \multirow{2}{*}{\cite{sleep_apnea1}} & \multirow{2}{*}{\cite{patient_monitoring}} & \multirow{2}{*}{\cite{cw_radar2}} & \multirow{2}{*}{\cite{cw_radar1}} & \multirow{2}{*}{\cite{uwb_radar2}} & \multirow{2}{*}{\cite{uwb_radar3}} & \multirow{2}{*}{\cite{uwb_radar4}} & \multirow{2}{*}{\cite{fmcw_vs_uwb}} & \multirow{2}{*}{\cite{fmcw_vs_uwb}} & \multirow{2}{*}{\cite{fmcw_radar1}} & \multirow{2}{*}{\cite{fmcw_radar3}} & \multirow{2}{*}{\cite{fmcw_radar4}} & \multirow{2}{*}{\cite{fmcw_radar5}} & \multirow{2}{*}{\cite{fmcw_radar2}} & \multirow{2}{*}{\cite{chest_disp_ref2}} \\ \\ \hline \rowcolor{Gray}
\textbf{Type of radar} & FMCW & FMCW & CW & UWB & CW & CW & UWB & UWB & UWB & UWB & FMCW & FMCW & FMCW & FMCW & FMCW & FMCW & FMCW \\
\textbf{Max no. of subjects} & 5 & 1 & 1 & 1 & 1 & 1 & 1 & 1 & 1 & 1 & 1 & 1 & 2 & 2 & 3 & 2 & 1\\\rowcolor{Gray}
\textbf{Max range reported (m)} &  6 & 6 & 1.5 & 3.5 & 1 & 2 & 0.8 & 0.55 & 2.76 & 2.5 & 2.5 & 2.14 & 0.5 & 0.8 & 2.6 & 4.3 & 1.5 \\ 
\textbf{Range resolution (m)} & 0.058 & 0.058 & - & - & - & - & - & - & 0.36 & - & - & 0.033 & - & 0.043 & - & 0.043 & 0.94 \\ \rowcolor{Gray}
\textbf{Angle resolution ($\degree$)} & 15 & - & - & - & - & - & - & - & - & - & - & - & 30 & 17.5 & - & 15 & -\\
\textbf{BR MAE (BRPM)} & 0.39 & 2.75 & 2.05 & - & - & - & - & - & - & 0.46 & 0.48 & 1.10 & 1.29 & - & 1.65 & - & -\\ \rowcolor{Gray}
\textbf{HR MAE (BPM)} & 3.64 & 4.75 & 5.06 & 2.29 & - & 0.46 & 1.56 & - & - & 1.56 & 1.56 & 18.40 & 4.60 & - & 4.00 & - & - \\ \hline
\end{tabular}}
\label{comparison_table}
\end{table*}


\subsubsection{Results}
\begin{figure}
    \centering
    \includegraphics[width=0.8\linewidth]{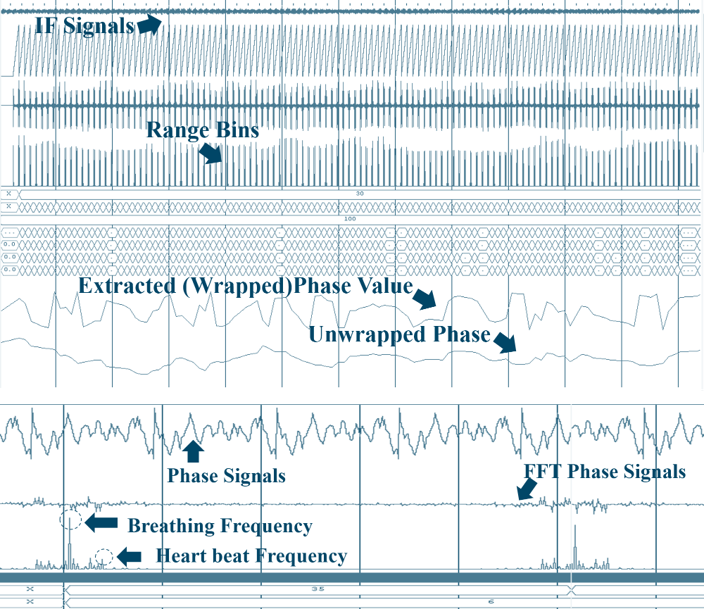}
    \caption{HDL simulation results for HR and BR estimation in Xilinx Vivado}
    \label{sim1}
\end{figure}

\begin{figure}
    \centering
    \includegraphics[width=1\linewidth]{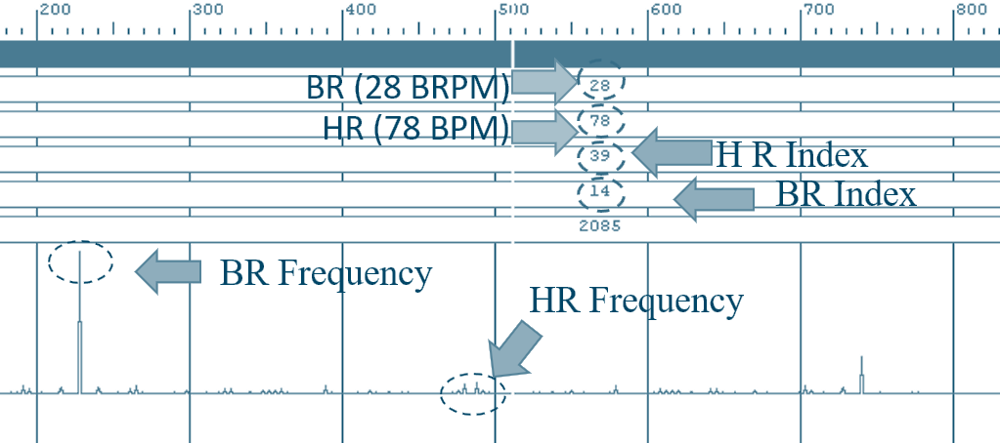}
    \caption{Hardware implementation results visualized on ILA}
    \label{ila}
\end{figure}

All the modules mentioned in Subsection \ref{subsub:FPGA_implementation} have been integrated to implement the proposed algorithm on ZedBoard FPGA. 
Fig. \ref{sim1} shows the HDL Simulation result for HR and BR estimation on Xilinx Vivado software.
Fig. \ref{ila} shows the implementation results visualized on \textcolor{Black}{ a} Integrated Logic analyzer (ILA).
Our extended investigation documented the power consumption outcomes, with reports showing a total dynamic power of 0.333W and a total on-chip power of 0.443W.
Table \ref{table: result table hr br} shows the results that demonstrate effective real-time HR/BR measurement for single subject scenarios.
We achieved much faster execution times for the ZedBoard FPGA based processing on 128 chirp data ($\approx$ 7402 total acceleration) than our software based processing on 512 chirp data. 
Table \ref{fpga resouce comaprarion} presents a comparative analysis of hardware resource usage and latency against existing literature for our FPGA-based processing. Our system shows superior performance in execution times and Look-Up Table (LUT) utilization, though with a slight increase in flip-flop (FF) and DSP block usage due to parallelism techniques.

\subsection{Comparison with Existing Works}
Table \ref{comparison_table} provides an extensive comparison of our work - both software and FPGA-based processing against prominent works in literature for endpoint results on HR/BR measurement performance. The HR/BR MAE of the other works that were not available was estimated using the accuracy reported and the mean HR/BR ground truth of our dataset. Our implementation achieves remarkable performance for estimating HR/BR, tolerable for up to subject distances of 6 m. Our BR MAE of 0.39 is the best among all other works. While our HR MAE is comparable with leading works, the superior BR MAE offers a balanced and advantageous trade-off.

\section{Conclusion} \label{sec:Conclusion}

This work proposes an algorithm with \textcolor{Black}{a} unique signal processing pipeline for simultaneous multi-subject HR and BR estimation using mmWave FMCW radar.
Unlike prior works, it overcomes the limitation of measuring vital signs from only one subject per azimuth. It offers detailed insights into trade-offs involving radar position, tilt, subject posture, range, and azimuth when the radar and subject’s chest 
are at different heights. The algorithm accurately measures HR and BR 
for up to five subjects simultaneously within a 6 m range, outperforming prior works.
The software implementation \textcolor{Black}{ using} MATLAB and Python achieved a best-in-class BR \textcolor{Black}{MAE} of 0.39, and HR MAE of 3.64. 
Additionally, we have developed an FPGA-based system, demonstrating a fully hardware-based, portable solution.
This FPGA implementation executes 2.7 times faster, uses 18.4\% fewer LUTs, and provides over 7400 times acceleration compared to its software counterpart, highlighting its potential as a superior alternative to traditional CPU or general-purpose solutions.


\section*{Acknowledgement}
The authors would like to acknowledge the Chips to Startup (C2S) program, Ministry of Electronics and Information Technology (MeitY), Govt. of India, 
IHub Mobility, IIIT Hyderabad, 
Kohli Center on Intelligent Systems (KCIS), IIIT Hyderabad,  
and IHub Anubhuti-IIIT Delhi Foundation for supporting this research.

\bibliographystyle{IEEEtran}
\bibliography{References_final}

\vspace*{-2.7\baselineskip}
\end{document}